\documentclass[superscriptaddress,twocolumn,showpacs,a4paper,
amssymb,amsmath,nobibnotes,aps,prd,
showkeys,
nofootinbib,notitlepage]{revtex4-1}
\usepackage{verbatim}
\usepackage[T1]{fontenc}
\usepackage[utf8]{inputenc}
\usepackage[american]{babel}
\usepackage{epsfig}
\usepackage{booktabs}
\usepackage{multirow}
\usepackage{dcolumn}
\usepackage{amsmath}
\usepackage{mathtools}
\usepackage{amsfonts}
\usepackage{amssymb}
\usepackage{ulem}
\usepackage{epstopdf}
\usepackage{bm}
\usepackage{siunitx}
\usepackage{braket}
\usepackage{enumitem}
\usepackage{soul}
\usepackage[table]{xcolor}
\usepackage{color}
\usepackage{transparent}
\usepackage{pifont}
\usepackage{enumitem}

\definecolor{navyblue}{rgb}{0.0, 0.0, 0.5}
\definecolor{royalblue}{rgb}{0.25, 0.41, 0.88}
\definecolor{cadmiumgreen}{rgb}{0.0, 0.42, 0.24}
\definecolor{blue-violet}{rgb}{0.54, 0.17, 0.89}
\definecolor{darkviolet}{rgb}{0.58, 0.0, 0.83}
\definecolor{orange(colorwheel)}{rgb}{1.0, 0.5, 0.0}

\usepackage{hyperref}
\hypersetup{
    colorlinks=true, 
    linkcolor=royalblue, 
    citecolor=magenta}

\usepackage{booktabs}
\usepackage{multirow}
\usepackage{dcolumn}
\usepackage{colortbl}

\begin{document}

\title{Measuring the reionization optical depth without large-scale CMB polarization}

\author{William Giar\`e}
\email{w.giare@sheffield.ac.uk}
\affiliation{School of Mathematics and Statistics, University of Sheffield, Hounsfield Road, Sheffield S3 7RH, United Kingdom}

\author{Eleonora Di Valentino}
\email{e.divalentino@sheffield.ac.uk}
\affiliation{School of Mathematics and Statistics, University of Sheffield, Hounsfield Road, Sheffield S3 7RH, United Kingdom}

\author{Alessandro Melchiorri}
\email{alessandro.melchiorri@roma1.infn.it}
\affiliation{Physics Department and INFN, Universit\`a di Roma ``La Sapienza'', Ple Aldo Moro 2, 00185, Rome, Italy}

\begin{abstract}
We study the possibility of measuring the optical depth at reionization, $\tau$, without relying on large-scale Cosmic Microwave Background (CMB) polarization. Our analysis is driven by the need to obtain competitive measurements that can validate the state-of-the-art constraints on this parameter, widely based on E-mode polarization measurements at $\ell\le 30$. This need is partially motivated by the typical concerns regarding anomalies observed in the Planck large-scale CMB data as well as by the remarkable fact that, excluding these latter, $\tau$ consistently exhibits correlations with anomalous parameters, such as $A_{\rm lens}$ and $\Omega_k$, suggesting that slightly higher values of the optical depth at reionization could significantly alleviate or even eliminate anomalies. Within the $\Lambda$CDM model, our most constraining result is $\tau = 0.080 \pm 0.012$, obtained by combining Planck temperature and polarization data at $\ell > 30$, the Atacama Cosmology Telescope (ACT) and Planck measurements of the lensing potential, Baryon Acoustic Oscillations (BAO), and Type-Ia supernova data from the Pantheon+ catalogue. Notably, using only ACT temperature, polarization, and lensing data in combination with BAO and supernovae, we obtain $\tau = 0.076 \pm 0.015$, which is entirely independent of Planck. The relative precision of these results is approaching the constraints based on large-scale CMB polarization ($\tau = 0.054 \pm 0.008$). Despite the overall agreement, we report a slight $1.8\sigma$ shift towards larger values of $\tau$. We also test how these results change by extending the cosmological model. While in many extensions they remain robust, in general obtaining precise measurements of $\tau$ may become significantly more challenging.
\end{abstract}

\date{\today}
\maketitle


\section{Introduction}

The epoch of reionization arguably stands as one of the central questions in modern cosmology. The quandary of when and how the first celestial objects in the Universe emitted a sufficient amount of ultraviolet radiation able to reionize the neutral hydrogen and helium within the intergalactic medium is an active area of research and debate~\cite{Vishniac:1987wm,Shapiro:1993hn,Hu:1993tc,Haiman:1996rc,Gnedin:1996qr,Miralda-Escude:1998adl,Knox:1998fp,Haiman:1996ht,Gnedin:1999fa,Shaver:1999gb,Hu:1999vq,Barkana:2000fd,Loeb:2000fc,Bullock:2000wn,Gnedin:2000uj,Fan:2001vx,Barkana:2001gr,Kaplinghat:2002vt,Wyithe:2002qu,Santos:2003pc,SDSS:2001tew,Hui:2003hn,Haiman:2003ea,Ciardi:2003ia,Malhotra:2004ef,Fan:2005es,McQuinn:2005hk,Iliev:2005sz,Kashikawa:2006pb,Bolton:2007fw,Mesinger:2007pd,Loeb:2008hg,Morales:2009gs,Kistler:2009mv,Parsons:2009in,Robertson:2010an,Zahn:2011vp,Pober:2013jna,Becker:2013ffa,Robertson:2013bq,Paciga:2013fj,Pentericci:2014nia,Becker:2014oga,Fialkov:2014kta,McGreer:2014qwa,Robertson:2015uda,Ali:2015uua,Madau:2015cga,Livermore:2016mbs,Patil:2017zqk,Davies:2018yfp,Park:2018ljd,Kulkarni:2018erh,Pagano:2019tci,Choudhury:2022rlm,Gnedin:2022eza,Sailer:2022vqx,Jain:2023jpy,Choudhury:2023fkm,Keating:2023jta,Mertens:2023dcl,Minoda:2023hvp,Ma:2023oko,Saxena:2023tue,Wolz:2023gql}. In this regard, the era of precision cosmology, made possible by a wide array of satellite and ground-based Cosmic Microwave Background (CMB) experiments such as WMAP~\cite{WMAP:2003pyh,WMAP:2003elm,WMAP:2006bqn,WMAP:2008rhx,WMAP:2008lyn,WMAP:2010qai,WMAP:2012nax,WMAP:2012fli}, Planck~\cite{Planck:2013pxb,Planck:2013win,Planck:2015fie,Planck:2015mrs,Planck:2015fie,Planck:2015mrs,Planck:2018vyg,Planck:2018nkj,Planck:2018lbu,Planck:2019nip} and more recently the Atacama cosmology telescope (ACT) ~\cite{ACT:2020gnv,ACT:2020frw,ACT:2023kun,ACT:2023dou} and the South Pole telescope (SPT)~\cite{SPT:2004qip,SPT-3G:2014dbx,SPT:2017jdf,SPT-3G:2021wgf,SPT-3G:2021eoc}, has undoubtedly marked a significant turning point.

In broad terms, during cosmic reionization, CMB photons undergo Thomson scattering off free electrons at scales smaller than the horizon size. As a result, they deviate from their original trajectories, reaching us from a direction different from the one set during recombination. Similarly to recombination, this introduces a novel 'last scattering' surface at later times and produces distinctive imprints in the angular power spectra of temperature and polarization anisotropies. A well-known effect of reionization is an enhancement of the spectrum of CMB polarization at large angular scales alongside a suppression of temperature anisotropies occurring at smaller scales. While the damping effect may overlap with variations in other cosmological parameters -- most prominently with the amplitude of primordial inflationary fluctuations $A_s$ -- the distinctive polarization bump produced by reionization on large scales dominates the signal in the $EE$ spectrum ($C_{\ell}^{EE}$) whose amplitude strongly depends on the total integrated optical depth to reionization:
\begin{equation}
\tau= \sigma_{\mathrm{T}} \int_0^{z_{\rm{rec} }} dz \, \bar n_{e}(z) \, \frac{dr}{dz},
\label{eq:reio}
\end{equation}
where $\sigma_{\mathrm{T}}$ is the Thomson scattering cross-section, $\bar n_{e}(z)$ is the free electron proper number density at redshift $z$ (with the over-line denoting an average over all sky directions), and $dr/dz$ is the line-of-sight proper distance per unit redshift. \footnote{Notice that the integral~\eqref{eq:reio} runs from recombination ($z_{\rm rec}\simeq 1100$) all the way up to today ($z=0$). However, setting an upper limit in the integral $z_{\rm max}\simeq 50$, is typically enough to capture the entirety of the expected contribution from reionization~\cite{Planck:2018vyg}.}

For this reason, precise observations of $E$-mode polarization on large scales are crucial at least for a dual purpose. On one side, they currently represent the most precise way to determine the value of $\tau$. Thanks to large-scale polarization measurements released by the Planck satellite, we have achieved an unprecedented level of accuracy, constraining the optical depth at reionization down to $\tau=0.054\pm0.008$~\cite{Planck:2018vyg} at $68\%$ confidence level (CL hereafter). On the other hand, measuring $\tau$ to such a level of precision holds implications that extend beyond reionization models. It is no exaggeration to say that it facilitates an overall refinement of cosmological constraints by alleviating many degeneracies resulting from similar effects on the spectra of temperature and polarization anisotropies caused by variations in different parameters. Just to mention a few remarkable examples supporting this last claim, we note that constraints on the Hubble parameter $H_0$ and the scalar spectral index $n_s$ both improve by approximately $22\%$ when incorporating Planck large-scale polarization data in the analysis.

However, despite the remarkable success of large-scale CMB polarization measurements -- which undoubtedly represent a significant achievement in modern cosmology and provide key insights into the physics of reionization -- as often happens when dealing with high-precision measurements at low multipoles, there are certain aspects that remain less than entirely clear:

\begin{itemize}

\item Firstly, the detected signal in the $EE$ spectrum is extremely small, at the order of $10^{-3} - 10^{-2} \, \mu K^2$~\cite{Pagano:2019tci}. On scales where cosmic variance sets itself a natural limit on the maximum precision achievable, this implies that current measurements are approaching the limits of experimental sensitivity, and even minor undetected systematic errors could have a substantial impact on the results. 

\item Secondly, the influence of galactic foregrounds, although significantly better understood in recent years, especially thanks to high-frequency measurements by the Planck satellite~\cite{Planck:2019nip}, remains more pronounced in relation to polarization anisotropies at equivalent angular scales. Consequently, small, undetected foreground effects could also play a role in determining polarization measurements~\cite{Luparello:2022kqb,Hansen:2023gra}.

\item Lastly, measurements of temperature and polarization anisotropies at large angular scales exhibit a series of anomalies. Around the lowest multipoles (corresponding to the largest scales) the $TT$ spectrum ($C_{\ell}^{\rm TT}$), deviates lower than expected within the best-fit cosmological model. Multipoles $\ell \lesssim 10$ (particularly the quadrupole and octopole modes) display unexpected features and correlations~\cite{Planck:2018vyg,Planck:2019nip}. Similarly, the $TE$ spectrum ($C_{\ell}^{\rm{TE}}$) show excess variance compared to simulations, most notably at $\ell=5$ and at $\ell=18 - 19$, for reasons that are not understood~\cite{Planck:2018vyg,Planck:2019nip}. As a result, these data are commonly disregarded for cosmological data analyses~\cite{Planck:2018vyg}. While little information is lost by discarding the $TE$ spectrum, it is still worth noting that, as for $C_{\ell}^{\rm{EE}}$, also $C_{\ell}^{\rm{TE}}$ is dominated by the optical depth at reionization on low multipoles; precisely the scales where unexplained anomalies are observed.
\end{itemize}

In light of these considerations, a few scattered questions and moderate concerns arise. Could the anomalies observed at low multipoles in $TT$ and $TE$ be present also in the $EE$ spectrum (at the very same scales) but remain undetected due to the smallness of the signal and the significant experimental noise? If so, can other anomalies typically encountered when extending the minimal $\Lambda$CDM cosmology (e.g., the lensing and curvature anomalies) somehow recast a wrong calibration of $\tau$?  More generally, is it possible to achieve \textit{competitive} constraints on $\tau$ without exclusively relying on large-scale CMB polarization?

In this paper, we aim to provide a comprehensive answer to all these points. We anticipate that, in fact, there exist other cosmic data that, when combined with small-scale temperature anisotropies, can provide valuable constraints on the optical depth even in the absence of large-scale polarization measurements. The ongoing advancements in our understanding of the large-scale structure of the Universe made possible by accurate reconstructions of the lensing potential~\cite{Planck:2018lbu,Carron:2022eyg,ACT:2023kun,ACT:2023dou}, Baryon Acoustic Oscillations measurements~\cite{BOSS:2012dmf,BOSS:2013rlg,BOSS:2014hwf,BOSS:2016wmc,BOSS:2013uda,eBOSS:2020yzd,SDSS:2003eyi,SDSS:2004kqt,SDSS:2006lmn,SDSS:2014iwm} and observations of Type-Ia supernovae~\cite{SupernovaSearchTeam:1998fmf,SupernovaCosmologyProject:1998vns,Pan-STARRS1:2017jku,Scolnic:2021amr,Brout:2022vxf}, are rapidly approaching a precision level that enables to narrow down the constraints on cosmological parameters, possibly breaking their degeneracy with $\tau$ on small scales. In light of these advancements, it is certainly timely to reevaluate the constraints on the optical depth obtained without large-scale CMB polarization and determine whether they align with (or diverge from) these latter.

The paper is structured as follows: in \autoref{sec:motivations}, we discuss in more detail the physical motivations in light of which we believe it is timely to achieve a measurement of $\tau$ independent from large-scale polarization. In \autoref{sec:Method}, we point out the methodology and data exploited through the analysis. In \autoref{sec:LCDM} and \autoref{sec:extensions}, we present the results obtained within the $\Lambda$CDM cosmology and its extensions, respectively. Finally, in \autoref{sec:Conclusion}, we derive our conclusion.

\section{Five Reasons why}
\label{sec:motivations}
From the discussion outlined in the introduction, several valid reasons have already emerged to conclude that obtaining a measurement of the optical depth at reionization independent from large-scale polarization can be an important -- if not even necessary -- step for both cross-checking the results and addressing some concerns involving anomalies in large-scale CMB data. However, in this section, we would like to further elaborate on the motivations that have led us to embark on the comprehensive analysis presented in this article. 

Before going any further, we want to clearly outline the terminology. First and foremost, we recall that the data released by the Planck satellite for the temperature-temperature spectrum $C_{\ell}^{TT}$, temperature-polarization spectrum $C_{\ell}^{TE}$, and the polarization-polarization spectrum $C_{\ell}^{EE}$ can be broadly categorized into two groups: low-multipole (large-scale) data within the range $2 \le \ell \le 30$ and high-multipole (small-scale) data at $\ell> 30$. In this study, we will use different combinations of these measurements resulting from different experiments or likelihoods. To ensure clear identification of the specific datasets and avoid any source of confusion, when dealing with the Planck temperature, polarization and lensing data we will consistently adopt the following nomenclature:
\begin{itemize}
\item \texttt{TT} refers to measurements of the power spectrum of temperature anisotropies $C_{\ell}^{TT}$ at small scales $\ell >30$ as obtained by the Planck likelihood \texttt{plik}~\cite{Planck:2018vyg,Planck:2019nip};

\item \texttt{TTTEEE} refers to measurements of the power spectra of temperature and polarization anisotropies $C_{\ell}^{TT}$,$C_{\ell}^{TE}$ and $C_{\ell}^{EE}$ at small scales $\ell >30$ as obtained by the Planck likelihood \texttt{plik}~\cite{Planck:2018vyg,Planck:2019nip};

\item \texttt{lowT} refers to measurements of the spectrum of temperature anisotropies $C_{\ell}^{TT}$ at large scales $2\le \ell \le 30$ as obtained by the Planck likelihood \texttt{Commander}~\cite{Planck:2018vyg,Planck:2019nip};

\item \texttt{lowE} refers to measurements of the spectrum of E-mode polarization $C_{\ell}^{EE}$ at large scales $2\le \ell \le 30$ as obtained by the Planck likelihood \texttt{SimAll}~\cite{Planck:2018vyg,Planck:2019nip};

\item \texttt{plik-lensing} refers to reconstruction of the spectrum of lensing potential (trispectrum) as obtained by the Planck collaboration~\cite{Planck:2018lbu};

\item \texttt{Planck-2018} refers to the full combination of all the above-mentioned data, namely \texttt{TTTEEE+lowT+lowE+plik-lensing}~\cite{Planck:2018vyg,Planck:2019nip}.
\end{itemize}

Notice that, for the reasons discussed in the introduction, we do not consider measurements of $C_{\ell}^{TE}$ at $\ell < 30$, as commonly done in the literature~\cite{Planck:2018vyg,Planck:2019nip}.

Keeping this nomenclature in mind, we now discuss in detail five different reasons why we find it particularly important to obtain measurements of $\tau$ that are independent of \texttt{lowE} (and possibly \texttt{lowT}) data.

\subsection{Consistency Test}
Let's start with the elephant in the room: acquiring independent measurements of the same parameter through different datasets represents one of the most effective and reliable methods to cross-check the results and ensure their validity. 

In this regard, it should be mentioned that the observational constraints on the reionization optical depth have changed quite a lot over time, due to a combination of a better understanding of foreground contamination and overall experimental improvements. 

Historically, one of the initial measurements of this parameter was derived from the first year of Wilkinson Microwave Anisotropy Probe (WMAP) observations, resulting in a constraint $\tau = 0.17 \pm 0.06$~\cite{WMAP:2003elm}. Within the same experiment, due to improvements in measurements and additional data, this result has undergone substantial changes. After three years of data collection, the value quoted by the WMAP collaboration in Ref.~\cite{WMAP:2006bqn} was $\tau = 0.089 \pm 0.030$. Subsequently, after five and seven years, they obtained $\tau = 0.084 \pm 0.016$~\cite{WMAP:2008lyn} and $\tau = 0.087 \pm 0.014$~\cite{WMAP:2010sfg}, respectively. Finally, the value quoted after the final maps and results from the WMAP nine-year observation was $\tau = 0.089 \pm 0.014$~\cite{WMAP:2012nax,WMAP:2012fli}.

The same aura of uncertainty has characterized the measurement of this parameter, even in the results provided by the Planck satellite experiment. Interestingly, the first value provided in the Planck-2013 results was $\tau = 0.089 \pm 0.032$, quoted in Tab.~II of Ref.~\cite{Planck:2013pxb}. The turning point, when the value of the optical depth started decreasing, was with the Planck-2015 results: $\tau = 0.066 \pm 0.016$~\cite{Planck:2015fie}. However, it is worth noting that, the same year, the Planck collaboration conducted the first detailed analysis of the Planck E-modes spectrum at high multipoles, quoting $\tau = 0.078 \pm 0.019$~\cite{Planck:2015bpv}. Moving to more recent times, in the Planck-2018 paper, we finally obtain the state-of-the-art constraint on this parameter, $\tau = 0.054 \pm 0.008$~\cite{Planck:2018vyg}. 

Despite a detailed analysis of the reasons behind the evolution of the constraints on this parameter is well beyond the scope of this article, it is now widely acknowledged that evidence of dust contamination in the WMAP large-scale polarization data ($2 < \ell < 23$ in the $TE$ spectrum) could have potentially impacted the early constraints on $\tau$. On the other hand, the improvements obtained by the Planck collaboration over the years are based on a better understanding of dust emission and foregrounds. The combination of these two facts motivated a global re-analysis of WMAP data using the Planck 353-GHz map as a dust template, leading to $\tau = 0.062 \pm 0.012$, see page 24 of Ref.~\cite{Planck:2019nip}. In any case, this brief investigation should be enough to highlight how, while perhaps never really being at the center of attention, accurately measuring the optical depth at reionization has been (and maybe still is) a significant challenge in precision cosmology. In light of this, we believe that obtaining accurate measurements, possibly based on diversified and non-contradictory data, could significantly contribute to validating the current results on this parameter. Moreover, such an approach could provide useful benchmark measurements for broader analyses. This is particularly relevant when analyzing current ground-based CMB data released from ACT and SPT, where Planck-based priors on $\tau$ are typically assumed~\cite{ACT:2020gnv,ACT:2023kun,ACT:2023dou,SPT-3G:2021wgf,SPT-3G:2021eoc}.

\subsection{Large-scale $E$-mode polarization measurements}

\begin{figure}
    \centering
    \footnotesize
    \includegraphics[width=\columnwidth]{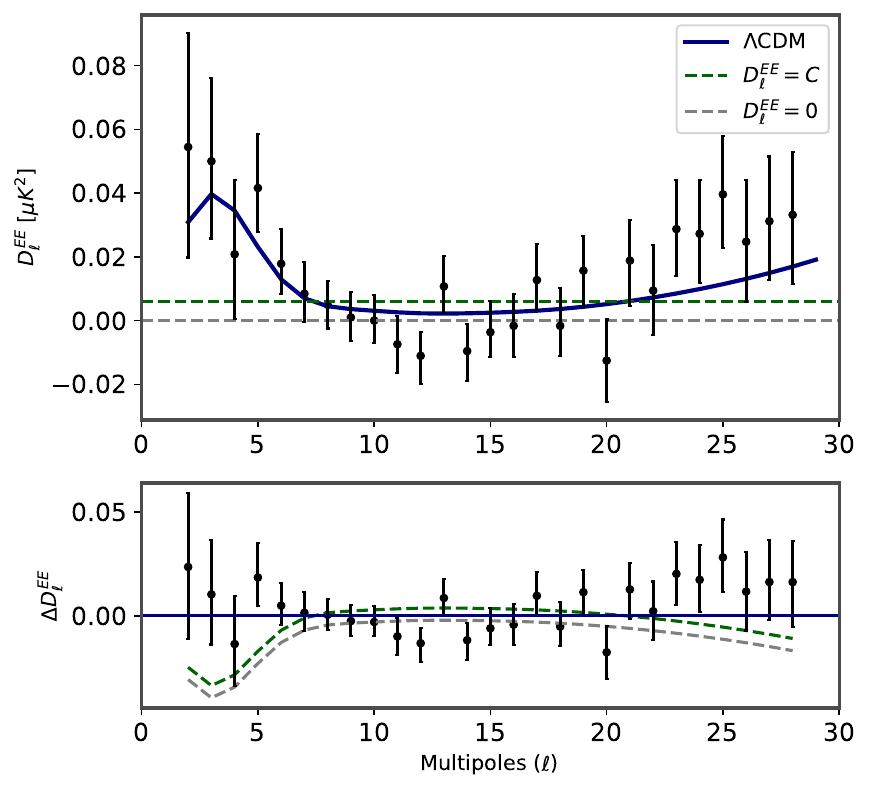}
    \includegraphics[width=0.945\columnwidth]{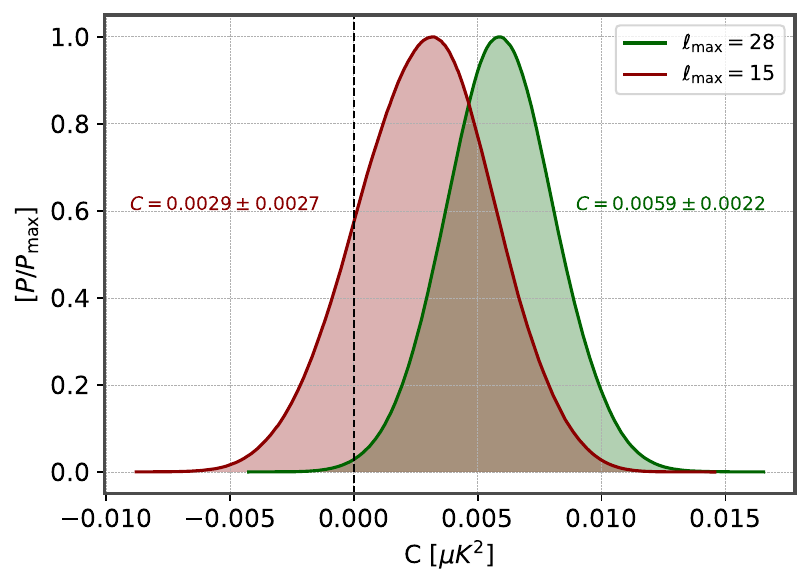}
    \caption{\textbf{Top Panel:} Theoretical predictions (and relative deviation in terms of residuals) for $\mathcal{D}_{\ell}^{EE}$ at $\ell<30$. The blue line represents the predictions derived within $\Lambda$CDM by fixing all cosmological parameters to their best-fit values based on the \texttt{Planck-2018} likelihoods. The green line represents the best-fit scenario for the toy-model case $\mathcal{D}{\ell}^{EE}=C$ . \textbf{Bottom Panel:} Posterior distribution function for $C$, considering the entire multipole range (in green) versus solely the first 15 multipoles (in red).}
    \label{fig:lowE}
\end{figure}

The most recent constraints on the reionization optical depth from the Planck satellite~\cite{Planck:2018vyg,Pagano:2019tci} are almost entirely based on measurements of the E-mode polarization at $\ell\le 30$. An easy exercise that can support this claim involves setting all the cosmological parameters to the best-fit values from \texttt{Planck-2018}, leaving only $\tau$ and $A_s$ free to be determined by data. Doing so, we obtain $\tau = 0.0508 \pm 0.0085$ and $\log(10^{10} A_s) = 3.032 \pm 0.019$ by using only \texttt{lowE} in combination with a gaussian prior $A_s e^{-2\tau} = \left(1.873\pm 0.016\right)\times 10^{-9}$. Alternatively, instead of including a prior on $A_s e^{-2\tau}$, one can directly combine \texttt{lowE+lowT} to get $\tau = 0.0521\pm 0.0086$ and $\log(10^{10} A_s) = 2.965\pm 0.052$, respectively. This unequivocally demonstrates the importance of temperature and polarization data at $\ell\le 30$. These measurements provide an excellent method for breaking the degeneracy between $A_s$ and $\tau$ appearing at smaller scales, allowing precise measurements of both parameters. It is also worth noting that these measurements are quite robust. The inclusion of additional parameters does not significantly alter the results. For instance repeating the same analysis by varying also the spectral index $n_s$ we obtain $\tau = 0.052^{+0.012}_{-0.013}$, $\log(10^{10} A_s) = 3.034\pm 0.025$, and $n_s = 0.977^{+0.069}_{-0.16}$ from \texttt{lowE} in combination with the aforementioned prior on $A_s e^{-2\tau}$, or $\tau = 0.0512\pm 0.0091$, $\log(10^{10} A_s) = 2.88^{+0.26}_{-0.29}$, and $n_s = 0.947^{+0.060}_{-0.12}$ from \texttt{lowE+lowT}.

Having established the importance of large-scale CMB measurements in determining $\tau$ and the robustness of the results obtained, we note that this situation represents both a blessing and a curse. On one hand, it has made it possible to determine $\tau$ within a relative precision of approximately 14\%. On the other hand, it is not an overstatement to say that relying so much on one dataset to measure a parameter that, as we will see, carries important implications for various issues emerged in recent years, can be at least imprudent.

This is even more true when one considers the difficulties surrounding precise CMB measurements at large angular scales, where foreground contamination have historically been a significant source of uncertainty. As mentioned in the introduction, measurements of temperature and polarization anisotropies at large angular scales exhibit a series of anomalies. In the \texttt{lowT} dataset, multipoles with $\ell < 10$ display unexpected features and correlations while, for reasons that are not understood, the $TE$ spectrum exhibits excess variance compared to simulations at low multipoles, where in principle the signal should be dominated by the optical depth at reionization. Fortunately, \texttt{lowE} appears to be anomaly-free since (as of now) there are no compelling arguments suggesting problems or unreliability in this dataset. However, it remains a fact that the signal in the spectrum of E-mode polarization at $\ell \le 30$ is extremely small, on the order of  $\sim 10^{-3} - 10^{-2} \mu K^2$. This implies that current measurements are approaching the limits of experimental sensitivity, and minor undetected systematic errors could potentially have a significant impact on the results.

To formalize this concern in a more quantitative way, we consider the most recent measurements of the spectrum of E-modes polarization in the multipole range $2\le \ell \le 30$ as provided in Ref.~\cite{Pagano:2019tci} (and shown in \autoref{fig:lowE}) and we figure out to what extent the signal is different from zero. To do so, we perform a fit to measurements of $\mathcal D_{\ell}^{EE}=\ell(\ell+1)C_{\ell}^{EE}/2\pi$ assuming a constant function $\mathcal D_{\ell}^{EE}=C$, where $C$ represents the amplitude of the spectrum in units of $\mu K^2$. We work under the unrealistic assumption that data points are independent and Gaussian distributed. While we know this is not the case and that the effects of non-Gaussianity and correlations among data are significant for large-scale EE measurements, we believe that this approximation is enough for the purpose of our toy-analysis. Considering the full multipole range $\mathcal D_{\ell}\in [2\le \ell \le 30]$ from the Markov Chains Monte Carlo (MCMC) analysis, we get $C=0.0059\pm0.0022 \, \mu K^2$ at 68\% CL, together with a best-fit value $C=0.00596 \, \mu K^2$. The $p$-value of the best fit $\mathcal D_{\ell}^{EE}=0.00596 \mu K^2$ is $p=0.063$. Despite all the important caveats surrounding this result, we note that the $p$-value remains above the threshold value typically adopted to reject the hypothesis ($p=0.05$). This is somewhat surprising since from the theoretical predictions for the spectrum of E-modes polarization in that multipole range one would expect $C_{\ell}^{EE}\propto \tau^2 / \ell^4$ and thus $\mathcal D_{\ell}^{EE}\propto \tau^2 / \ell^2$. However, when considering the full multipole range $C_{\ell}^{EE}\in [2\le \ell \le 30]$, the $p$-value for the case $\mathcal D_{\ell}^{EE}=0$ is $p=0.012$; well below the threshold value of $p=0.05$. That being said, looking at \autoref{fig:lowE}, the signal appears to substantially diminish on scales $4\lesssim \ell \lesssim 15$. In fact, 8 out of the first 15 data points are consistent with $D_{\ell}^{EE} = 0$ within $1\sigma$. These scale are those that contribute more when determining $\tau$ because it is where the characteristic reionization bump in polarization manifests itself more prominently. Therefore we repeat the analysis focusing only on data-points at $2\le\ell\le15$. In this case we obtain $C=0.0029\pm0.0027\,\mu K^2$ with a best-fit value of $0.00287\,\mu K^2$. As we shall see from the posterior distribution function in the bottom panel of \autoref{fig:lowE}, the case $C=0$ (i.e., no signal at all) falls basically within the 1$\sigma$ range. In addition, the $p$-value for the best fit scenario, $\mathcal{D}_{\ell}^{EE}=0.00287\,\mu K^2$, is $p=0.087$, while the $p$-value corresponding to $\mathcal{D}_{\ell}^{EE}=0\,\mu K^2$ is $p=0.064$. Both of them are larger than the threshold used for dismissing the hypothesis.

In conclusion, when examining the residuals reported in \autoref{fig:lowE}, it is evident that while $\Lambda$CDM provides a good fit to these measurements (better than the hypothesis of a null or constant signal), the data points in \texttt{lowE} remain highly correlated, and the broad trends in residuals are not significant enough to conclusively confirm a reionization bump. As our toy-analysis has shown, a constant or vanishing signal can, in principle, be fitted to this dataset without being ruled out by the p-value statistics. In addition, we acknowledge the concern that, when dealing with measurements so close to the absence of a signal and experimental sensitivity, any statistical fluctuation or lack of understanding of the foreground could be crucial and potentially have implications in the measurement of $\tau$. Certainly, an independent measurement of this parameter would provide reassurance, dispelling any doubts about using large-scale polarization measurements.

\subsection{Lensing Anomaly}

One of the standout achievements of the standard $\Lambda$CDM model of cosmology is its ability to provide a robust fit to the Planck data. Nonetheless, despite this undeniable success, in recent years a few mild anomalies have emerged, becoming the subject of intense study. Among them, one issue that has been at the heart of discussions and keeps stimulating significant research interest~\cite{DiValentino:2015bja,Renzi:2017cbg,Domenech:2020qay,Addison:2023fqc,Kable:2020hcw,Addison:2015wyg} is the higher lensing amplitude observed in the temperature and polarization spectra.

\begin{table}[htbp!]
\begin{center}
\renewcommand{\arraystretch}{2}
\resizebox{0.9 \columnwidth}{!}{
\begin{tabular}{l | c | c }
\hline
\textbf{Planck Likelihood} & \boldmath{$\tau$} & \boldmath{$A_{\rm lens}$}\\ 
\hline\hline

\texttt{TT} & $< 0.178$ & $1.13\pm 0.13$ \\

\texttt{TT}+\texttt{plik-lensing} & $< 0.175$ & $0.99\pm 0.10$\\

\texttt{TT}+\texttt{lowT} & $< 0.123$ & $1.23\pm 0.12$ \\

\texttt{TT}+\texttt{lowT}+\texttt{plik-lensing} & $< 0.119$ & $1.075\pm 0.079$ \\

\texttt{TT}+\texttt{lowE} & $0.0499\pm 0.0085$ & $1.205\pm 0.099$ \\

\texttt{TT}+\texttt{lowE}+\texttt{plik-lensing} & $0.0494\pm 0.0086$ & $1.058\pm 0.054$\\

\texttt{TT}+\texttt{lowT}+\texttt{LowE} & $0.0500\pm 0.0087$ & $1.243\pm 0.096$ \\

\texttt{TT}+\texttt{lowT}+\texttt{LowE}+\texttt{plik-lensing} & $0.0496\pm 0.0084$ & $1.082\pm 0.052$ \\

\hline
\texttt{TTTEEE}  & $< 0.168$ & $1.09\pm 0.12$ \\

\texttt{TTTEEE}+\texttt{plik-lensing} & $< 0.171$ & $0.987\pm 0.096$\\

\texttt{TTTEEE}+\texttt{lowT} & $< 0.115$ & $1.174\pm 0.095$ \\

\texttt{TTTEEE}+\texttt{lowT}+\texttt{plik-lensing} & $<0.114$ & $1.065^{+0.082}_{-0.074}$\\

\texttt{TTTEEE}+\texttt{lowE} & $0.0495\pm 0.0086$ & $1.168\pm 0.066$ \\

\texttt{TTTEEE}+\texttt{lowE}+\texttt{plik-lensing} & $ 0.0497\pm 0.0086$ & $1.061\pm 0.042$\\

\texttt{TTTEEE}+\texttt{lowT}+\texttt{LowE} & $0.0492\pm 0.0086$ & $1.180\pm 0.065$ \\

\texttt{Planck-2018} & $0.0491\pm 0.0084$ & $1.071\pm 0.040$ \\

\hline \hline
\end{tabular}}
\end{center}
\caption{Constraints on $\tau$ and $A_{\rm lens}$ obtained within $\Lambda$CDM+$A_{\rm lens}$ for all possible combinations of the Planck likelihoods. Constraints are given at 68\% while upper bounds are given at 95\% CL.}
\label{tab.Alens}
\end{table}

\begin{figure}[htbp!]
    \centering
    \includegraphics[width=0.65\columnwidth]{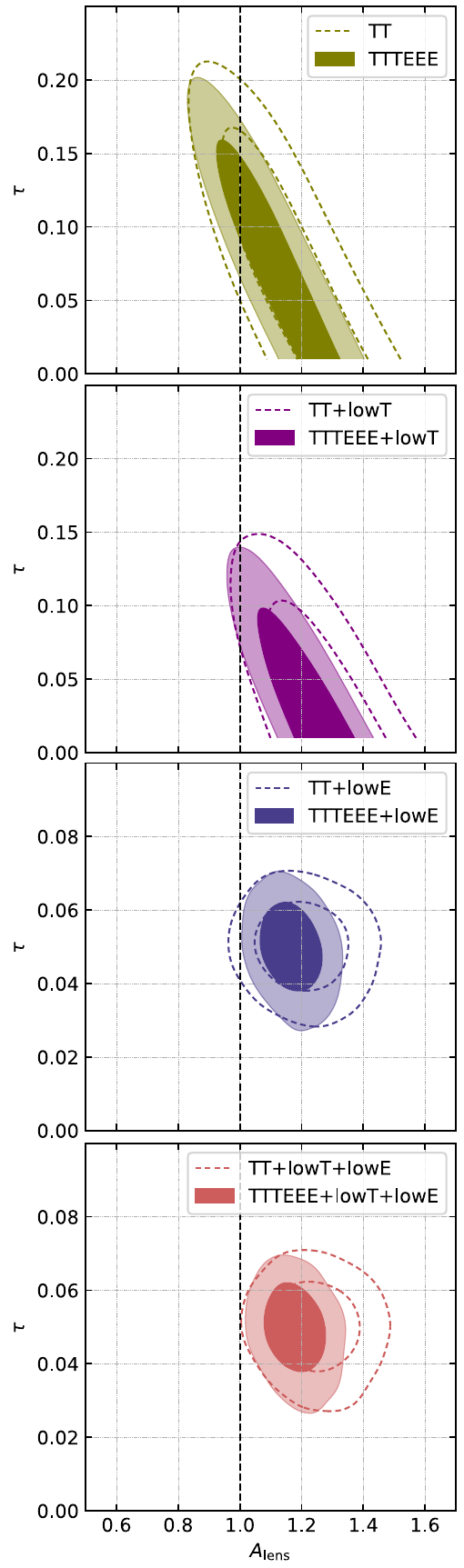}
    \caption{Joint marginalized contours at 68\% and 95\% CL illustrating the correlation between the lensing amplitude $A_{\rm lens}$ and the optical depth at reionization $\tau$ for different combinations of the Planck likelihoods.}
    \label{fig:Alens}
\end{figure}

To put it in more quantitative terms, the lensing anomaly can be quantified by the phenomenological parameter $A_{\rm lens}$ -- first introduced in Ref.~\cite{Calabrese:2008rt} --  that captures any deviations from the lensing amplitude expected within $\Lambda$CDM, corresponding to $A_{\rm lens}=1$. As well-known, apart from the direct measurement of the CMB lensing spectrum, the lensing amplitude manifests in the temperature and polarization power spectra through the lensing-induced smoothing of the acoustic peaks and the transfer of power to the damping tail. Surprisingly, when $A_{\rm lens}$ is treated as a free parameter, the analysis of Planck~\texttt{TTTEEE+lowE+lowT} data yields $A_{\rm lens}=1.180 \pm 0.065$, suggesting an excess of lensing in the spectra of temperature and polarization anisotropies at about $2.8\sigma$~\cite{Planck:2018vyg}.

The Planck preference for $A_{\rm lens} > 1$ is not due to a volume effect in the parameter space, but it results from a genuine improvement in the $\chi^2$ by approximately $\Delta \chi^2 \sim 9.7$~\cite{Planck:2018vyg}. Analysis of the temperature and polarization spectra has widely established that this improvement primarily comes from \texttt{TTTEEE} data, particularly within the multipole range $600 < \ell < 1500$. Nevertheless, allowing $A_{\rm lens}$ to vary also improves the fit to \texttt{lowT} data. Regarding \texttt{TT} data, one can even visually observe a preference for increased lensing smoothing in the oscillatory residuals at $1100 < \ell < 2000$, which matches the shape of the lensing smoothing (see, e.g., Fig 24 in \cite{Planck:2018vyg}).

Within $\Lambda$CDM, the fact that \texttt{TTTEEE} data prefer more lensing translates into a preference for higher fluctuation amplitudes. Hence, the high-$\ell$ data typically yield higher values of $A_s$ and $\tau$ than large-scale E-mode polarization. This suggests from the onset a strong negative correlation between $\tau$ and $A_{\rm lens}$: higher values of the former can modulate the effects of a lensing amplitude $A_{\rm lens} > 1$. Clearly, in the presence of \texttt{lowE} data, such a correlation is immediately broken as $\tau$ is measured down to 14\% precision. However, considering our earlier concerns about large-scale polarization data, one might question whether the lensing anomaly could be, at least in part, produced by a misalignment in the calibration of the signal within the \texttt{lowE} data. Given its proximity to a case with no signal at all, any fluctuations could artificially lower the values of $\tau$. In other terms, relaxing the constraints on reionization, one would naively expect to recover agreement in $A_{\rm lens}$ by shifting up $\tau$. Not surprisingly, this is exactly what we find when analyzing all possible combinations of the Planck Likelihoods within the $\Lambda$CDM+$A_{\rm lens}$ framework. The results are summarized in \autoref{tab.Alens} and can be visually represented in \autoref{fig:Alens}.

In the plot, each panel represents constraints in the plane ($A_{\rm lens}$ , $\tau$) obtained through various combinations of Planck data. Let's start with the bottom panel in the figure. We can see that by combining small and large scale measurements, the correlation between the two parameters is significantly reduced. Once again, this is because the \texttt{lowE} data precisely constrain $\tau$. Moving on to the second panel from the bottom, we see that the influence of \texttt{lowT} data, while not critical in this context, is still relevant: when these data are removed from the analysis, the negative correlation between the two parameters becomes slightly more pronounced. On the other hand, advancing to the next panel, we notice that excluding the \texttt{lowE} data significantly affects the contours, confirming our naive expectation that values of $A_{\rm lens}$ around 1 can be reintroduced by allowing for higher values of $\tau$. However, it's important to emphasize that $A_{\rm lens}=1$ is only marginally consistent with the 2-$\sigma$ contours and would still require unreasonably high values $\tau \sim 0.1$. Finally, let’s move to the top panel of the figure. We observe that in the absence of information on both temperature and polarization at large angular scales, a significant correlation between the two parameters is introduced. In this case, values of $A_{\rm lens}\sim 1$ become fully consistent with both \texttt{TT} and \texttt{TTTEEEE}. It's crucial to note that in this scenario, achieving $A_{\rm lens}=1$ does not require artificially high values of $\tau$. In fact, $\tau \sim 0.07 - 0.08$ is perfectly compatible with $A_{\rm lens}=1$.

In conclusion, an artificially low measurement of $\tau$ derived from large-scale (temperature and) polarization data may impact the lensing anomaly, potentially biasing the preference for $A_{\rm lens}>1$ observed in the Planck data. As we will demonstrate shortly, a similar argument can be extended to other anomalies, such as the curvature anomaly and the overall preference towards phantom dark energy. Therefore, obtaining competitive measurements of optical depth at reionization independently of large-scale polarization data holds intrinsic significance, as they are needed to either validate or highlight differences among results obtained with or without \texttt{lowE} data.


\subsection{Curvature Anomaly}

\begin{figure*}[htbp!]
    \centering
    \includegraphics[width=0.55\textwidth]{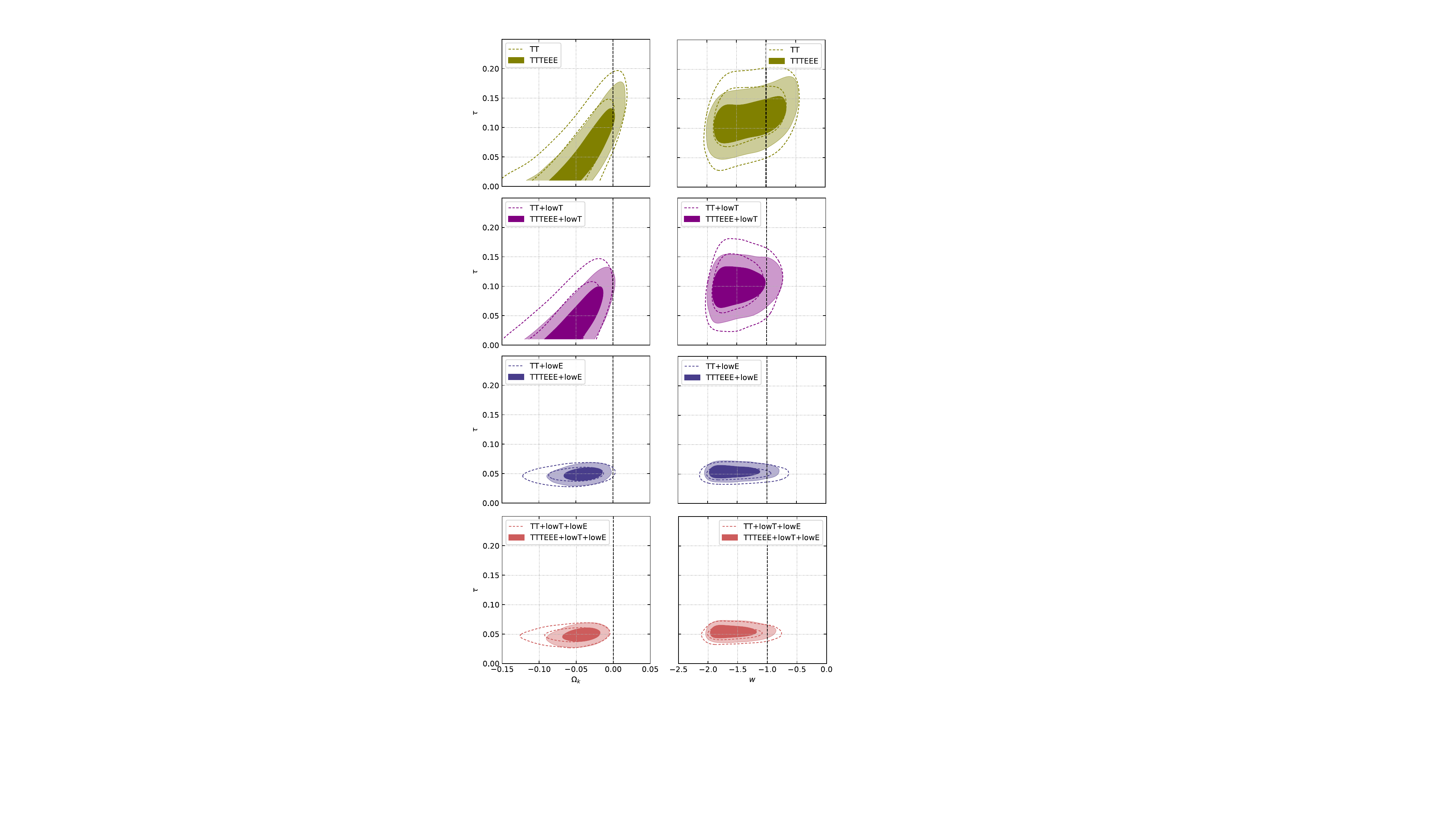}
    \caption{\textbf{Left panels:} Joint marginalized contours at 68\% and 95\% CL illustrating the correlation between the curvature density parameter $\Omega_k$ and the optical depth at reionization $\tau$ for different combinations of the Planck likelihoods. \textbf{Right panels:} Joint marginalized contours for the DE equation of state $w$ and $\tau$ for the same likelihoods.}
    \label{fig:omk_w}
\end{figure*}

The presence of lensing anomaly can have significant consequences, especially when it comes to constraining cosmological parameters beyond the standard $\Lambda$CDM model. A point typically cited to make this argument involves noting that, since more lensing is expected with a higher abundance of dark matter, the observed lensing anomaly can be recast into a preference for a closed Universe,\footnote{Notice, however, that the potential implications of the lensing anomalies extend far beyond the curvature parameter. Other notable examples concern modified gravity and the neutrino sector, without claiming to be exhaustive, see e.g.~\cite{DiValentino:2015bja,Pogosian:2021mcs,DiValentino:2021imh,Specogna:2023nkq,Nguyen:2023fip} and references therein.} well documented and discussed in several recent works~\cite{Park:2017xbl,Handley:2019tkm,DiValentino:2019qzk,Efstathiou:2020wem,DiValentino:2020hov,Benisty:2020otr,Vagnozzi:2020rcz,Vagnozzi:2020dfn,DiValentino:2020kpf,Yang:2021hxg,Cao:2021ldv,Dhawan:2021mel,Dinda:2021ffa,Gonzalez:2021ojp,Akarsu:2021max,Cao:2022ugh,Glanville:2022xes,Bel:2022iuf,Yang:2022kho,Stevens:2022evv,Favale:2023lnp}. Here we note that the analysis conducted on the lensing anomaly has shown that removing large-scale temperature and polarization data leads to a correlation between $\tau$ and $A_{\rm lens}$, such that slightly larger values of the former can possibly reconcile the lensing amplitude inferred from Planck temperature and polarization spectra at $\ell > 30$ with the expected value in $\Lambda$CDM. Therefore, given the strong interconnection between $A_{\rm lens}$ and $\Omega_k$, it is worth considering whether similar conclusions can also be drawn for the curvature anomaly.

To address this question, we can refer to the left panels in \autoref{fig:omk_w}, where two-dimensional correlations between $\tau$ and $\Omega_k$ are presented for different combinations of the Planck likelihoods. Starting from the bottom-left panel, we observe that constraints on these two parameters obtained, including both \texttt{lowE} and \texttt{lowT} data, exclude $\Omega_k=0$ at more than 95\% CL. Instead, from the second and third left panels from the bottom, we see that excluding respectively \texttt{lowE} and \texttt{lowT}, the global preference towards $\Omega_k < 0$ becomes weaker than the one obtained by including temperature anisotropies or E-mode polarization measurements at large scales. Excluding both \texttt{lowE} and \texttt{lowT} (upper left panel in the figure), we recover agreement with $\Omega_k=0$ well within the 68\% CL contours. Therefore, we can draw two significant conclusions. First, we can confirm that the part of the preference for $\Omega_k<0$ is due to the inclusion of \texttt{lowT} data. As it is well-known, they exhibit a deficit of power at large angular scales (most prominently at the quadrupole and octupole modes) that remains largely unexplained within the flat $\Lambda$CDM model and drives the shift towards $\Omega_k<0$. Second, we notice that $\tau$ appears to be correlated with $\Omega_k$ in a way such that slightly larger values of the optical depth at reionization lead to a reduced preference for a closed Universe. Moving to the upper-left panel, we can see that by removing large-scale temperature data, this result is not only confirmed but further accentuated. Now $\Omega_k = 0$ falls within two standard deviations for values of $\tau$ in the range of $\tau \sim 0.7 - 0.8$. Again, this stems from the fact that by relaxing the constraints on the optical depth to reionization, the excess lensing in the damping tail appears to recast into a preference for a larger $\tau$, similar to what happens with $A_{\rm lens}$.

\subsection{Shift towards phantom Dark Energy}
The last of the five reasons we want to discuss to underscore the importance of obtaining an independent measurement of $\tau$ pertains, one more time, unexpected results derived from the Planck data when extending the cosmological model. In this case, we focus on the Dark Energy (DE) equation of state parameter $w$. The constraints on this parameter has recently undergone a comprehensive reanalysis in light of the most updated data~\cite{Escamilla:2023oce}. As extensively discussed in Ref.~\cite{Escamilla:2023oce}, by considering \texttt{Planck-2018}, one can find that at $68\%$ ($95\%$) CL $w = -1.57^{+0.16}_{-0.36}$ ($w = -1.57^{+0.50}_{-0.40}$) confirming the consistent findings documented in the literature~\cite{Planck:2018vyg, Yang:2021flj}, indicating a Planck preference for a phantom equation of state at a statistical level slightly exceeding two standard deviations. Such a  preference can be attributed to a wide range of physical and geometrical effects. Notably, when excluding the \texttt{lowE} likelihood, a significant shift of $w$ towards the cosmological constant value $w = -1$ is observed, and no preference for a phantom equation of state is found anymore.

This result can be further understood from the panels on the right in \autoref{fig:omk_w}, which display the two-dimensional correlations between $w$ and $\tau$. As visible in all panels in the figure, removing \texttt{lowE} leads to all other combinations of Planck data being consistent with $w = -1$ for values of $\tau \sim 0.07-0.08$.
\bigskip

{\textit{In conclusion, whether it's the lensing anomaly, the curvature anomaly, or the shift towards phantom DE, including or excluding large-scale CMB data (and hence determining or not determining $\tau$ through \texttt{lowE} and possibly \texttt{lowT}) produces a substantial difference. $\tau$ consistently displays correlations with anomalous parameters, suggesting that slightly higher values of the optical depth at reionization could go a long way in alleviating or even eliminating all these anomalies. As the saying goes, “three clues make a proof” and we can certainly use this proof to underscore the importance of deriving independent measurement of this parameter.}

\section{Data and Methods}
\label{sec:Method}
In light of the reasons documented in the previous two sections, we now proceed with our aim to derive a measure of $\tau$ that is both competitive and independent of large-scale CMB temperature and polarization measurements. To achieve this goal, we will examine a wide range of different datasets in various combinations. Specifically, we shall consider:

\begin{itemize}
    \item CMB measurements of the power spectra of temperature and polarization anisotropies $C_{\ell}^{TT}$, $C_{\ell}^{TE}$, and $C_{\ell}^{EE}$ at small scales $\ell > 30$ as obtained by the Planck likelihood \texttt{plik}~\cite{Planck:2018vyg,Planck:2019nip}. As already pointed out in \autoref{sec:motivations}, we refer to these datasets as \texttt{TT} when they include only temperature anisotropies and \texttt{TTTEEE} when they include polarization, as well.

    \item CMB measurements of the power spectra of temperature and polarization anisotropies $C_{\ell}^{TT}$, $C_{\ell}^{TE}$, and $C_{\ell}^{EE}$ at small scales $\ell > 600$ as obtained by the Atacama Cosmology Telescope ACT-DR4 likelihood~\cite{ACT:2020gnv,ACT:2020frw}. We will refer to this dataset as \texttt{ACT-DR4}.

    \item The gravitational lensing mass map, which covers 9400 deg$^2$, reconstructed from CMB measurements obtained by the Atacama Cosmology Telescope from 2017 to 2021~\cite{ACT:2023kun,ACT:2023dou}. We consider only the conservative range of lensing multipoles  $40 < \ell < 763$. This dataset is denoted as \texttt{ACT-DR6}.

    \item \texttt{ACT-DR6} is considered both independently and in conjunction with the Planck satellite experiment. Following Refs.~\cite{ACT:2023kun,ACT:2023dou}, when combining two likelihoods together, we use the more recent \texttt{NPIPE} data release~\cite{Carron:2022eyg} that re-processed Planck time-ordered data with several improvements, including around 8\% more data compared to the \texttt{plik-lensing} likelihood. We note that the \texttt{NPIPE} lensing map covers CMB angular scales in the range $100 \le \ell \le 2048$ using the quadratic estimator. Since \texttt{NPIPE} and \texttt{ACT-DR6} measurements overlap only across a portion of the sky, explore distinct angular scales, and exhibit varying noise levels and instrument-related systematics, they can be regarded as nearly independent lensing measurements. We refer to the final combination of \texttt{ACT-DR6} and \texttt{NPIPE} simply as \texttt{lensing}. This dataset should not be confused with the one denoted as \texttt{plik-lensing} in \autoref{sec:motivations}. Please, read the last sentence again.

    \item Observations of the local Universe in the form of:\\
    \\
    1) Baryon Acoustic Oscillation (BAO) and Redshift-Space Distortions (RSD) measurements from the completed SDSS-IV eBOSS survey. These include isotropic and anisotropic distance and expansion rate measurements, as well as measurements of $f\sigma_8$, and are summarized in Table~3 of Ref.~\cite{eBOSS:2020yzd}.\\
    \\
    2) 1701 light curves for 1550 distinct SNeIa in the redshift range $0.001 < z < 2.26$ collected in the \textit{PantheonPlus} sample~\cite{Scolnic:2021amr}. In all but one case, we will consider the uncalibrated \textit{PantheonPlus} SNeIa sample. \\
\\
We collectively denote these two datasets as \texttt{low-z}.
    
\end{itemize}

We perform the MCMC analyses employing the samplers \texttt{Cobaya}~\cite{Torrado:2020dgo} in conjunction with the Boltzmann solver \texttt{CAMB}~\cite{Lewis:1999bs}. To test the convergence of the chains obtained using this approach, we utilize the Gelman-Rubin criterion~\cite{Gelman:1992zz}, and we establish a threshold for chain convergence of $R-1 < 0.01$. 

\begin{table}[t]
	\begin{center}
		\renewcommand{\arraystretch}{1.5}
		\begin{tabular}{l@{\hspace{0. cm}}@{\hspace{3 cm}} c}
			\hline
			\textbf{Parameter}    & \textbf{Prior} \\
			\hline\hline
                $\tau$                       & $[0.001\,,\,0.8]$ \\
			$\Omega_{\rm b} h^2$         & $[0.005\,,\,0.1]$ \\
			$\Omega_{\rm c} h^2$         & $[0.001\,,\,0.99]$ \\
			$100\,\theta_{\rm {MC}}$     & $[0.5\,,\,10]$ \\
			$\log(10^{10}A_{\rm s})$     & $[1.6\,,\,3.9]$ \\
			$n_{\rm s}$                  & $[0.8\,,\, 1.2]$ \\
   			$\sum m_{\nu}$ [eV]          & $[0\,,\,5]$\\
                $N_{\rm eff}$     	         & $[0.05\,,\,10]$\\
			$\Omega_{\rm k} $     	     & $[-0.3\,,\,0.3]$\\
			$w_0$                        & $[-3\,,\,1]$ \\
			$w_a$                        & $[-3\,,\,2]$ \\
                $\alpha_{\rm s}$             & $[-1\,,\, 1]$ \\
                $A_{\rm lens}$               & $[0\,,\, 10]$ \\
			\hline\hline
		\end{tabular}
		\caption{List of uniform prior distributions for cosmological parameters.}
		\label{tab.Priors}
	\end{center}
\end{table}

As for the cosmological model, since the key point of our analysis is to derive \textit{robust} bounds on the optical depth at reionization without large-scale CMB data, a necessary step is to ensure that the results remain stable when extending the background cosmology. Therefore, while in \autoref{sec:LCDM} we start considering the baseline $\Lambda$CDM scenario, in \autoref{sec:extensions} we eventually consider a plethora of possible minimal extended cosmologies. As a result, along with the six $\Lambda$CDM parameters (i.e., the amplitude $A_s$ and the spectral index $n_s$ of scalar perturbations, the baryon $\Omega_b h^2$ and the cold dark matter $\Omega_c h^2$ energy densities, the angular size of the sound horizon at recombination $\theta_{\rm MC}$ and the reionization optical depth, $\tau$), we will also consider additional parameters such as the sum of neutrino masses $\sum m_\nu$, the number of relativistic degrees of freedom $N_{\textrm{eff}}$, the running of the scalar index $\alpha_s$, the curvature component $\Omega_k$, the DE equation of state parameters $w$ and $w_0$-$w_a$, and the lensing amplitude $A_{\rm lens}$. We refer to \autoref{tab.Priors} for the priors adopted in all the cosmological parameters.

\section{Results for $\boldmath{\Lambda}$CDM}
\label{sec:LCDM}

\begin{table*}[htbp!]
\begin{center}
\renewcommand{\arraystretch}{1.8}
\resizebox{\textwidth}{!}{
\begin{tabular}{l c c c c c c c c c c c c c c c }
\hline
\textbf{Parameter} & \textbf{ 	\texttt{TT+lensing} }  & \textbf{ 	\texttt{TT+lensing+low-z} } & \textbf{ 	\texttt{TTTEEE+lensing} } & \textbf{ 	\texttt{TTTEEE+lensing+low-z} } & \textbf{ \texttt{ACT(DR4+DR6)} } & \textbf{ \texttt{ACT(DR4+DR6)+low-z}} \\ 
\hline\hline

$ \tau_\mathrm{reio}  $ & $  0.078\pm 0.025\, ( 0.078^{+0.049}_{-0.049} ) $ & $  0.079\pm 0.013\, ( 0.079^{+0.025}_{-0.024} ) $ & $  0.078\pm 0.016\, ( 0.078^{+0.033}_{-0.032} ) $ & $  0.080\pm 0.012\, ( 0.080^{+0.023}_{-0.023} ) $ & $  0.094^{+0.037}_{-0.048}\, ( 0.094^{+0.071}_{-0.082} ) $ & $  0.076\pm 0.015\, ( 0.076^{+0.030}_{-0.031} ) $ \\ 
$ \Omega_\mathrm{b} h^2  $ & $  0.02230\pm 0.00027 $ & $  0.02230\pm 0.00020 $ & $  0.02249\pm 0.00017 $ & $  0.02250\pm 0.00014 $ & $  0.02162\pm 0.00030 $ & $  0.02162\pm 0.00029 $ \\ 
$ \Omega_\mathrm{c} h^2  $ & $  0.1184\pm 0.0027 $ & $  0.1184\pm 0.0011 $ & $  0.1187\pm 0.0016 $ & $  0.11857\pm 0.00098 $ & $  0.1167\pm 0.0043 $ & $  0.1188\pm 0.0013 $ \\ 
$ 100\theta_\mathrm{MC}  $ & $  1.04103\pm 0.00053 $ & $  1.04103\pm 0.00041 $ & $  1.04105\pm 0.00032 $ & $  1.04106\pm 0.00029 $ & $  1.04232\pm 0.00074 $ & $  1.04210\pm 0.00062 $ \\ 
$ n_\mathrm{s}  $ & $  0.9678\pm 0.0080 $ & $  0.9681\pm 0.0045 $ & $  0.9683\pm 0.0053 $ & $  0.9687\pm 0.0040 $ & $  1.005\pm 0.016 $ & $  0.999\pm 0.012 $ \\ 
$ \log(10^{10} A_\mathrm{s})  $ & $  3.086\pm 0.043 $ & $  3.088\pm 0.022 $ & $  3.087\pm 0.029 $ & $  3.091\pm 0.021 $ & $  3.110\pm 0.065 $ & $  3.081\pm 0.026 $ \\ 
\hline
$ H_0  $ & $  67.9\pm 1.2 $ & $  67.90\pm 0.52 $ & $  67.96\pm 0.72 $ & $  68.00\pm 0.44 $ & $  68.4\pm 1.8 $ & $  67.54\pm 0.51 $ \\ 
$ \sigma_8  $ & $  0.824\pm 0.011 $ & $  0.8247\pm 0.0076 $ & $  0.8247\pm 0.0088 $ & $  0.8260\pm 0.0073 $ & $  0.844\pm 0.017 $ & $  0.837\pm 0.010 $ \\   
\hline \hline
\end{tabular} }
\end{center}
\caption{ Constraints at 68\% (95\%) CL on cosmological parameters obtained within the $\Lambda$CDM model of cosmology by different datasets.}
\label{tab.results.lcdm}
\end{table*}

We start by analyzing the constraints that can be derived on $\tau$ without large-scale CMB data, assuming a standard $\Lambda$CDM cosmology. In \autoref{tab.results.lcdm} we summarize all the most important results obtained by considering various combinations of data, ranging from CMB observations at small angular scales to observations of the local Universe at low redshift. The same results can be visualized in \autoref{fig:tau.lcdm}.

\subsection{Reionization Optical Depth}

First and foremost, we consider the combination of data denoted as \texttt{TT+lensing}. It is interesting to note that precise lensing measurements released by Planck and ACT, when combined with the observations of temperature anisotropies on small scales, lead to a constraint on the optical depth to reionization, $\tau = 0.078 \pm 0.025$ ($0.078^{+0.049}_{-0.049}$) at the 68\% (95\%) CL. In this case, $\tau$ is measured with a relative precision of about 32\%. Therefore, this constraint is not competitive with the one obtained using the full \texttt{Planck-2018} combination ($\tau = 0.0543 \pm 0.0076$), where the optical depth is constrained up to 14\% precision. Nonetheless, this result relies on a relatively limited amount of data, making it evident that it is indeed plausible to obtain constraints on reionization without large-scale CMB data as well as that precise lensing measurements are crucial for this purpose.

As a second step, we consider the same combination of data adding \texttt{low-z} information about the local Universe (i.e., BAO and SNIa measurements). This leads to an improvement in constraints on the reionization optical depth, yielding a value $\tau = 0.079 \pm 0.013$ $(0.079^{+0.025}_{-0.024})$ at the 68\% (95\%) CL. In this case, the relative precision increases up to 16.5\%, approaching the level achieved by \texttt{lowE} data. Such an improvement further underscores the crucial role played by low-redshift information. As extensively documented in the literature, these datasets are essential for breaking correlations among cosmological parameters induced by similar effects on the spectrum of temperature anisotropies\footnote{Without large-scale polarization, we observe a loss of constraining power on $\tau$ and $A_s$. This introduces additional degeneracy lines. For instance, varying $\Omega_c h^2$ primarily alters the amplitude of all acoustic peaks, and higher values of $A_s$ can be compensated by lower values of $\Omega_c h^2$. By adding \texttt{low-z} data, we can accurately determine $\Omega_m$, which primarily contributes to fixing $\Omega_c h^2$, breaking the degeneracy with $A_s$ and allowing consequently more precise measurements of $\tau$.}. It is worth noting that the value obtained for $\tau$, although in good agreement with \texttt{Planck-2018} within two standard deviations, shows a shift towards higher values.

We proceed by including information on polarization measurements at small scales. Analogously to the previous cases, we start with \texttt{TTTEEE}+\texttt{lensing}. We obtain $\tau=0.078\pm 0.016$ ($0.078^{+0.033}_{-0.032}$) at the 68\% (95\%) CL. Incorporating polarization data significantly improves the constraint obtained from \texttt{TT}+\texttt{lensing}, allowing us to achieve approximately 20\% precision. That being said, the precision remains lower compared to \texttt{TT}+\texttt{lensing}+\texttt{low-z}. It's also worth noting that, even though all results are broadly consistent within two standard deviations, the trend towards higher values of $\tau$ compared to the full \texttt{Planck-2018} combination is confirmed also when including polarization measurements.

\begin{figure}[htpb!]
    \centering
    \includegraphics[width=\columnwidth]{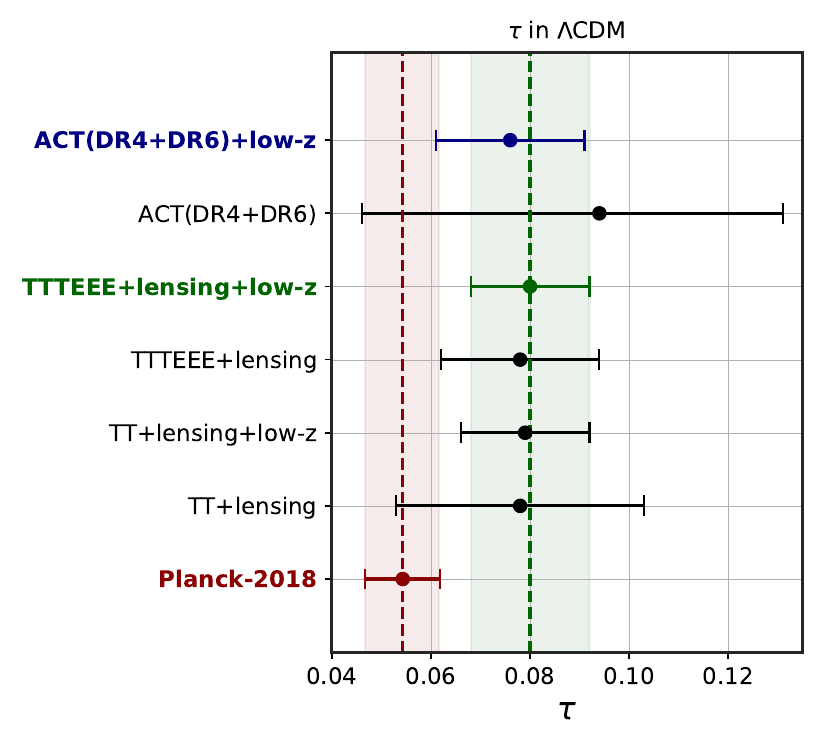}
    \caption{Values and corresponding 1$\sigma$ errors of $\tau$ obtained through different combinations of data within $\Lambda$CDM. \texttt{Planck-2018} (in red) is the only dataset containing temperature and polarization measurements at large angular scales. \texttt{TTTEEE+lensing+low-z} (in green) provides the most stringent constraint among the \texttt{lowE}-free datasets considered in this study, serving as our consensus dataset. \texttt{ACT(DR4+DR6)+low-z} (in blue) provides a Planck-independent measurement of this parameter.}
    \label{fig:tau.lcdm}
\end{figure}

By including the \texttt{low-z} data and considering the combination \texttt{TTTEEE+lensing+low-z}, we obtain what, to the best of our knowledge, represents the most precise constraint on the reionization optical depth in the absence of large-scale temperature and polarization data. Specifically, we find $\tau=0.080\pm 0.012$ ($0.080^{+0.023}_{-0.023}$) at 68\% (95\%) CL. In this case, the parameter $\tau$ is constrained with a relative precision of 15\%, and this precision becomes comparable to that obtained when considering the \texttt{lowE} data. Therefore, we consider this result significant since the good accuracy reached in measuring $\tau$ can serve as an important test for the standard results obtained by \texttt{Planck-2018}. In particular, we note that, although both measurements of $\tau$ are consistent within two standard deviations, the shift towards higher values in the absence of large-scale polarization data is confirmed and slightly increased. For this combination of data -- which represents the most stringent one considered in this study -- the shift reaches a statistical significance of $1.8\sigma$\footnote{The $1.8\sigma$ shift has been calculated by adopting the so-called "rule of thumb difference in mean", which involves comparing the difference in means of $\tau$ for two different datasets to the quadrature sum of the uncertainties, namely Eq.(40) in Ref.~\cite{Raveri:2018wln}. This procedure requires marginalizing over many parameters, possibly introducing volume effects and potentially washing out tensions or signals in the $D$-dimensional parameter space. For this reason, we have tested the global parameter shift in $D=6$ dimensions by adopting the statistical methodology introduced in Section VII of Ref.~\cite{Raveri:2019gdp}, which is valid for correlated datasets. Considering different combinations of datasets, we find that the global shift remains always comparable to the shift in $\tau$, ensuring that no sensible information is lost.}. Clearly, a shift of $1.8\sigma$ is certainly not enough to represent an element of concern, but it is interesting to note that this result appears to align with those obtained in previous years (or decades!) by experiments like WMAP and/or "early" Planck. The crucial difference among these "early" measurements and ours is that the latter does not depend on large-scale CMB data, where foreground contamination has always been a source of uncertainty. In fact, it relies only on solid and very well understood data. We also note that, as discussed in \hyperref[sec.AppendixB]{Appendix B}, the very same shift is also confirmed by analyzing small-scale Planck data from the newly released Planck likelihoods \texttt{CamSpec}~\cite{Rosenberg:2022sdy} and \texttt{HiLLiPoP}~\cite{Tristram:2023haj}, which come with several improvements compared to \texttt{plik}, above all accounting for more sky area at high frequencies and addressing several enhancements in the processing of time-ordered data and foreground modeling.
That being said, it remains true that large part of these data comprises the temperature, polarization, and lensing spectra measured by Planck. Therefore, another crucial step is to obtain a precise measurement of $\tau$ that is entirely independent of Planck.

To this end, we consider measurements of temperature and polarization anisotropies released by \texttt{ACT-DR4} in conjunction with the recent \texttt{ACT-DR6} lensing measurements provided by the same collaboration. We consider ACT data both on their own and in combination with \texttt{low-z} observations. Considering \texttt{ACT(DR4+DR6)} alone, we obtain $\tau=0.094^{+0.037}_{-0.048}$ ($0.094^{+0.071}_{-0.082}$) at the 68\% (95\%) CL. In this case, the uncertainties are certainly too large to draw reliable conclusions. However, when including low-redshift information, from \texttt{ACT(DR4+DR6)+low-z} we get $\tau=0.076\pm0.015$ ($0.076^{+0.030}_{-0.031}$), i.e., a measurement of $\tau$ with an accuracy of approximately 20\%, entirely independent of Planck. This measurement aligns perfectly with \texttt{TTTEEE+lensing+low-z}, confirming the trend towards slightly higher values of the reionization optical depth.

\begin{figure*}[htbp!]
    \centering
    \includegraphics[width=\textwidth]{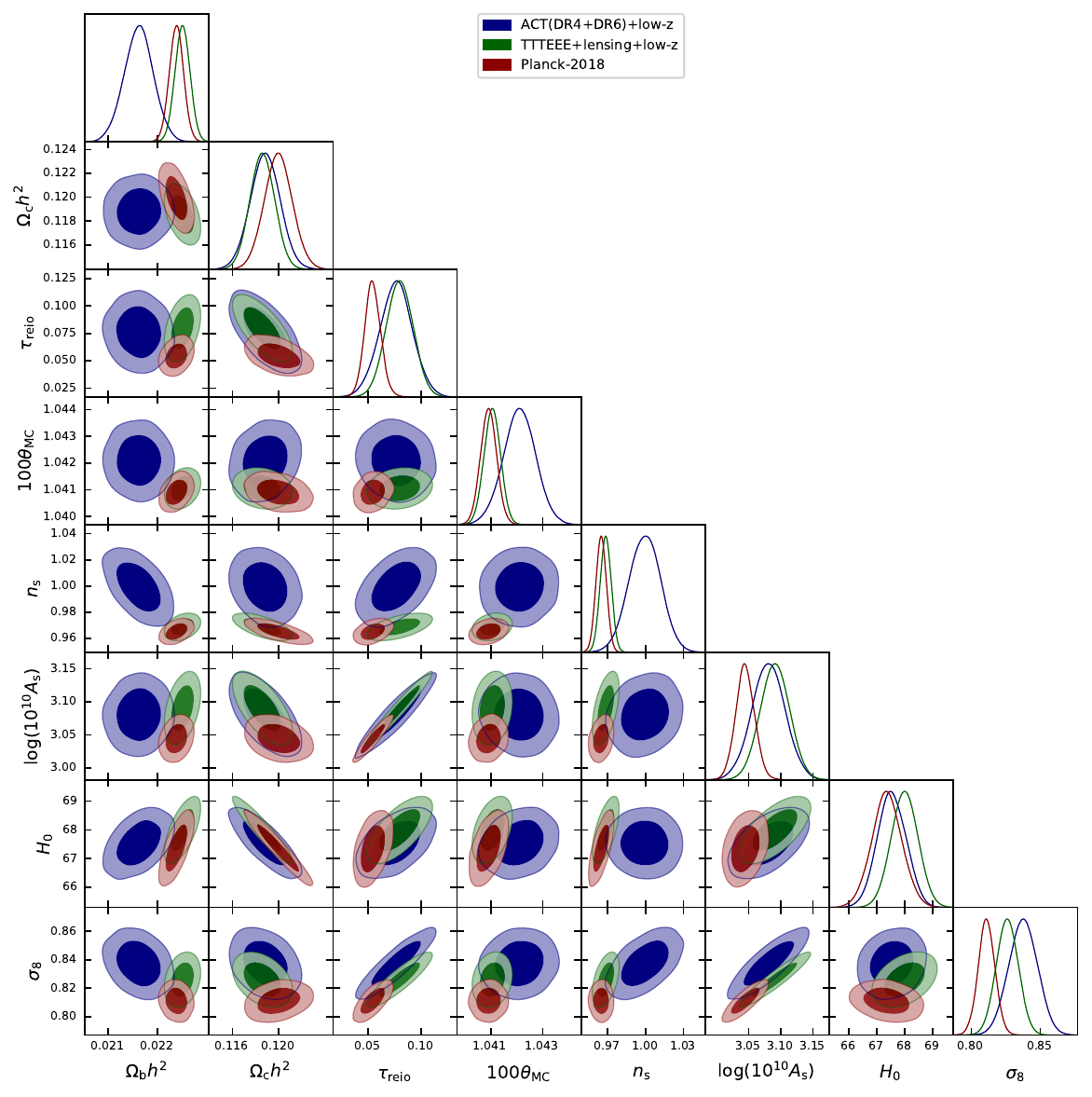}
    \caption{One-dimensional posterior probability distributions and two-dimensional contours at 68\% and 95\% CL for the $\Lambda$CDM parameters derived from the three reference datasets indicated in the legend.}
    \label{fig:tau.lcdm_Tplot}
\end{figure*}

We conclude this section taking a last look at \autoref{fig:tau.lcdm} where all the results discussed so far are summarized. In the figure, we identify three specific combinations of data that we consider particularly noteworthy. First and foremost, we consider the full \texttt{Planck-2018} case (shown in red in the figure) as the baseline case for comparison. Secondly, we identify \texttt{TTTEEE+lensing+low-z} (green in the figure) as our consensus dataset for measuring $\tau$ in the absence of large-scale CMB observations. We choose this combination as the consensus dataset because it provides the most constraining result and aligns well with the other \texttt{lowE}-free data combinations while incorporating a larger amount of information. Lastly, we identify \texttt{ACT(DR4+DR6)+low-z} (blue in the figure) as our "safety check" dataset. While less constraining than our consensus dataset, this combination of data is entirely Planck-independent and precise enough to double-check all the results we will mention in the subsequent sections.

\subsection{Implications for Cosmological Parameters}

The analysis detailed in the previous subsection confirmed that it is indeed possible to obtain measurements of $\tau$ independent of large-scale CMB polarization with a precision that is competitive and comparable to the latter. Furthermore, we report that in the absence \texttt{lowE} and \texttt{lowT} data, observations show a preference for higher values of the optical depth at reionization, leading to a shift in the results whose statistical significance reaches the level of 1.8 standard deviations for the consensus datasets (\texttt{TTTEEE+lensing+low-z}).

This shift, of course, has implications for the constraints obtained on other cosmological parameters. In \autoref{fig:tau.lcdm_Tplot}, for the three reference datasets identified at the end of the previous subsection, we show the one-dimensional marginalized posterior distribution functions and two-dimensional correlations for all parameters involved in the analysis. Leaving aside the well-documented shift~\cite{Lin:2019zdn,Handley:2020hdp,LaPosta:2022llv,DiValentino:2022oon,DiValentino:2022rdg,Giare:2022rvg,Giare:2023wzl,Giare:2023xoc,Calderon:2023obf} observed between the results based on Planck (red and green contours in the figure) and those based on ACT (blue contours in the figure) our analysis confirms a global preference of small-scale temperature and polarization measurements towards larger values of the amplitude of primordial inflationary fluctuations $A_s$. For our consensus dataset, the constraint on this parameter reads $\log\left(10^{10}\,A_s\right)=3.091\pm 0.021$, 1.8$\sigma$ higher compared to the the result obtained by \texttt{Planck-2018} ($\log\left(10^{10}\,A_s\right)=3.044\pm 0.015$). The shift in $A_s$ (which recast the one in $\tau$) can be compensated by a corresponding preference for a lower abundance of cold dark matter in the Universe. This is clear from the strong negative correlation between $\Omega_c h^2$ and $A_{\rm s}$ in the green contours in \autoref{fig:tau.lcdm_Tplot} and from the small shift in the final results ($\Omega_c h^2 = 0.11857\pm 0.00098$ from \texttt{TTTEEE+lensing+low-z} instead of $\Omega_c h^2 =0.1200\pm 0.0012$ from \texttt{Planck-2018}).

All these differences, and most prominently the increased amplitude of primordial perturbations, produce higher values of the parameter $\sigma_8$ which, at 68\% CL, reads $\sigma_8=0.8260\pm 0.0073$. This value is $1.6\sigma$ larger when compared to the value inferred by \texttt{Planck-2018} of $\sigma_8=0.8111\pm 0.0061$. As a result, excluding \texttt{lowE} and \texttt{lowT} data seems to further exacerbate the well-known tension between Weak Lensing (WL) and CMB experiments regarding the values of parameters governing the structure formation in the Universe~\cite{DiValentino:2020vvd}, although very recently the actual disagreement between these experiments appears to be the subject of careful reevaluation, see, e.g., Ref.~\cite{Kilo-DegreeSurvey:2023gfr}.

On the contrary, excluding \texttt{lowE} and \texttt{lowT} from the analysis does not appear to significantly reduce the ongoing tension between Planck~\cite{Planck:2018vyg} and SH0ES~\cite{Riess:2021jrx} regarding the value of the Hubble constant. From the consensus dataset, we obtain $H_0=68.00\pm0.44$ km/s/Mpc, very slightly shifted in the direction of local measurements. However, such a shift is not substantial enough to suggest any involvement of large-scale temperature and polarization measurements in the Hubble tension.

\section{Results for extensions to $\boldmath{\Lambda}$CDM}
\label{sec:extensions}

\begin{table}[tp!]
\begin{center}
\renewcommand{\arraystretch}{1.8}
\resizebox{\columnwidth}{!}{
\begin{tabular}{l | c | c | c }
 \hline
& \multicolumn{3}{c}{\textbf{Optical depth at reionization $\boldmath{\tau}$}}\\
\hline
\textbf{Model}  & \textbf{ 	\texttt{Planck-2018} } & \textbf{ 	\texttt{TTTEEE+lensing+low-z} } & \textbf{ \texttt{ACT(DR4+DR6)+low-z}} \\ 
\hline\hline
$\Lambda$CDM & $0.0543\pm 0.0076$ & $0.080\pm0.012$ & $0.076\pm 0.015$ \\

$\Lambda$CDM+$\sum m_{\nu}$  & $0.0553 \pm 0.0075$ & $0.095 \pm 0.016$ & $0.133 \pm 0.034$ \\

$\Lambda$CDM+$N_{\rm eff}$ & $0.0533\pm 0.0074$ & $0.080\pm0.012$ & $0.086 \pm 0.017$ \\

$\Lambda$CDM+$\alpha_s$ & $0.0553\pm 0.0077$ & $0.079 \pm 0.012$ & $0.055^{+0.018}_{-0.020}$ \\

$\Lambda$CDM+$\Omega_k$  & $0.0493\pm 0.0084$ & $0.079 \pm 0.013$ & $0.084 \pm 0.022$ \\

$w$CDM & $0.0524\pm 0.0074$ & $0.090 \pm 0.014$ & $0.098 \pm 0.022$ \\

$w_0 w_a$CDM  & $0.0521\pm 0.0075$ & $0.074 \pm 0.017$ & $0.076^{+0.028}_{-0.036}$ \\

$\Lambda$CDM+$A_{\rm lens}$ & $0.0491^{+0.0084}_{-0.0074}$ & $0.099 \pm 0.034$ & $0.133 \pm 0.034$\\
\hline \hline
\end{tabular} }
\end{center}
\caption{Constraints at 68\% CL on $\tau$ obtained from \texttt{Planck-2018}, our consensus dataset \texttt{TTTEEE+lensing+low-z}, and the Planck-independent dataset \texttt{ACT(DR4+DR6)+low-z} across various extended cosmologies.}
\label{tab.results.extensions}
\end{table}

The results discussed in the previous section prove that combining data on temperature and polarization anisotropies at small angular scales with precise reconstructions of the lensing potential spectrum and low-redshift information, it is possible to obtain constraints on the optical depth at reionization that are competitive with those derived by E-mode polarization measurements at large angular scales.

However, an important assumption upon which these constraints are derived is assuming a baseline $\Lambda$CDM model of cosmology. While this framework has undeniably excelled in explaining many observations, recent tensions are raising the question of whether it represents the end of the story or whether we need to introduce some new physics to restore concordance in cosmological and astrophysical observations. Just to mention a few concrete examples, the ongoing tension concerning the value of the Hubble constant between direct and indirect measurements could potentially reveal a mismatch in our understanding of the early and late-time Universe so that several theoretical attempts are trying to find a resolution by introducing new physics in the model either at early or late times (or even both, see e.g., Refs~\cite{Verde:2019ivm,DiValentino:2020zio,Perivolaropoulos:2021jda,DiValentino:2021izs,Abdalla:2022yfr,Hu:2023jqc,Vagnozzi:2023nrq} for reviews and discussions). Furthermore, in addition to the Hubble tension, a few other scattered and less significant anomalies exist, such as the already mentioned controversy surrounding the values of the matter cluster parameters ($S_8$ and $\sigma_8$)~\cite{DiValentino:2020vvd}, and recent measurements conducted by the James Webb Space Telescope~\cite{Gardner:2006ky} (JWST) indicating a higher density of massive galaxies at high redshift than previously assumed~\cite{2023ApJ...942L..27S,2022ApJ...938L..15C,2023ApJ...946L..13F,Treu:2022iti,2023ApJS..265....5H,2023ApJ...946L..16P,2023ApJ...949L..18P,Boylan-Kolchin:2022kae,Biagetti:2022ode}.\footnote{Particularly relevant for this article is the fact that JWST measurements show an overall disagreement with Planck polarization measurements~\cite{Forconi:2023izg}.}

\begin{figure}[tp]
    \centering
    \includegraphics[width=\columnwidth]{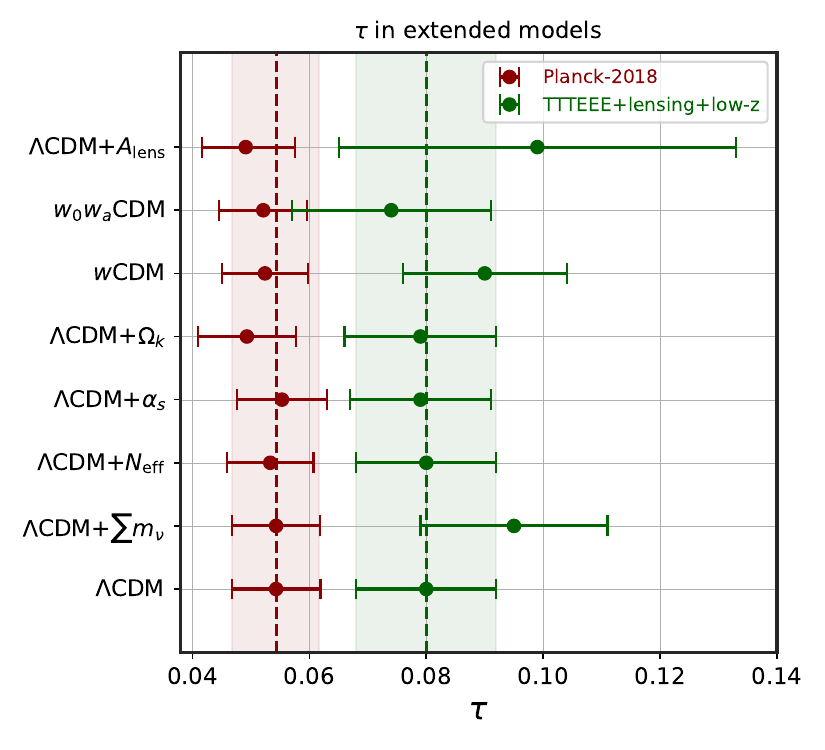}
    \caption{Values and corresponding 1$\sigma$ errors of $\tau$ obtained from \texttt{Planck-2018} (in red) and our consensus dataset \texttt{TTTEEE+lensing+low-z} (in green) in different extended models of cosmology.}
    \label{fig:tau.extensions}
\end{figure}

Even though it is certainly premature to draw definitive conclusions from these observations, it is not implausible that, in the near future, cosmology could undergo a new paradigm shift. Therefore, we believe it is imperative to understand to what extent the results derived in the previous section rely on assuming $\Lambda$CDM in the data analysis, as well as how they change beyond the standard cosmological model.

For this reason, we repeat the same analysis performed for the baseline case, considering the usual array of extended cosmologies typically analyzed when deriving constraints beyond $\Lambda$CDM~\cite{DiValentino:2015ola,DiValentino:2016hlg,DiValentino:2019dzu,DiValentino:2020hov,SPT-3G:2021wgf,Yang:2022kho,DES:2022ccp}. These models include additional parameters featuring new physics at both early (prior to recombination) and late (post-recombination) times. We keep in mind that the overall goal of this section is to study the implications for reionization -- specifically the results on the optical depth $\tau$ -- rather than testing new physics. That being said, we will, however, study the implications for beyond-$\Lambda$CDM parameters.

A summary of the results obtained in extended models are given in \autoref{tab.results.extensions}. Given the large amount of data and models considered, in this table we report only the results on $\tau$ derived for the consensus dataset (\texttt{TTTEEE+lensing+low-z}), testing them against both \texttt{Planck-2018} and the Planck-independent combination \texttt{ACT(DR4+DR6)+low-z}. However, in \hyperref[appendix:tables]{Appendix A} we provide tables with the full results for all the other data combinations and parameters considered in this study. We can quickly visualize the results in \autoref{tab.results.extensions} by referring to the \autoref{fig:tau.extensions} and \autoref{fig:tau.correlation}. In the first figure, we present constraints on the optical depth at reionization in models beyond $\Lambda$CDM, comparing results obtained from the consensus dataset with those from the \texttt{Planck-2108}. Meanwhile, in \autoref{fig:tau.correlation}, we illustrate the two-dimension correlation between $\tau$ and the additional beyond-$\Lambda$CDM parameters as obtained by the consensus dataset and \texttt{ACT(DR4+DR6)+low-z}.

\begin{figure*}[tp]
    \centering
    \includegraphics[width=\textwidth]{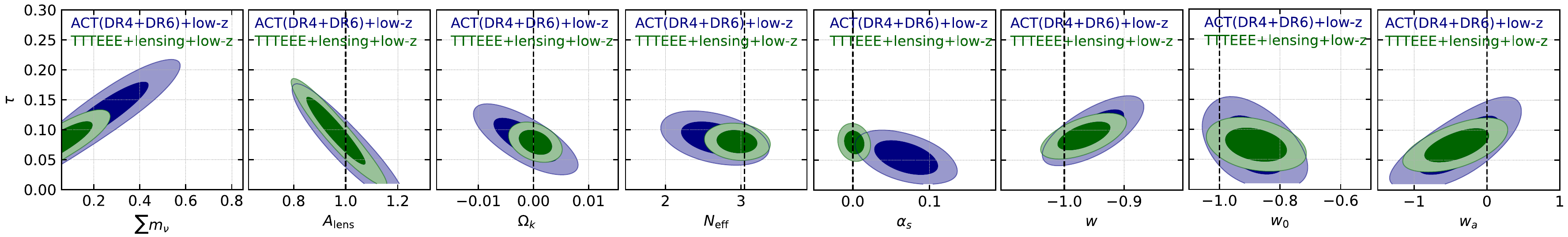}
    \caption{Two-dimensional correlations between $\tau$ and beyond-$\Lambda$CDM parameters.}
    \label{fig:tau.correlation}
\end{figure*}

\subsection{Massive Neutrinos}
We start our investigation into beyond-$\Lambda$CDM models by considering neutrinos as massive particles and leaving their total mass $\sum m_{\nu}$ as a free parameter to be constrained by data.

In this extension, from our consensus dataset \texttt{TTTEEE+lensing+low-z}, we obtain $\tau = 0.095 \pm 0.016$. This result can be compared to that obtained from \texttt{Planck-2018}, which reads $\tau = 0.0553 \pm 0.0075$, and with the Planck-independent combination \texttt{ACT(DR4+DR6)+low-z}, which places a constraint $\tau = 0.133 \pm 0.034$. Including the mass of neutrinos as an additional parameter leads to a general increase in uncertainties in determining $\tau$ as well as a further shift towards higher values. This is due to a strong positive correlation between these two parameters, as highlighted in \autoref{fig:tau.correlation}. This correlation is particularly pronounced in the \texttt{ACT(DR4+DR6)+low-z} as the temperature and polarization data released by ACT only cover angular scales $\ell \gtrsim 650$, thus lacking data around the first acoustic peak in the spectrum of temperature anisotropies, which are crucial for partially reducing the correlation between $\tau$ and $\sum m_{\nu}$.

Regarding the value of the neutrino mass, from the consensus dataset, we obtain $\sum m_{\nu} < 0.228$ eV, while considering \texttt{ACT(DR4+DR6)+low-z}, we get $\sum m_{\nu} < 0.476$ eV (both at a 95\% CL). Therefore, despite the correlation with $\tau$, the total neutrino mass is still well-constrained, primarily due to the addition of information at \texttt{low-z} and precise lensing spectrum measurements, which have both proven to be crucial in determining the properties of eV-scale thermal relics\footnote{Notice that late-time-only constraints on the total neutrino mass can be even stronger than early-time constraints in certain extended cosmologies~\cite{RoyChoudhury:2018gay,RoyChoudhury:2019hls,DiValentino:2021imh,DiValentino:2022oon,DiValentino:2022njd,Forconi:2023akg}. }~\cite{Giare:2023aix}.

Regarding other cosmological parameters, a closer inspection of \autoref{tab.results.lcdm_mnu} in \hyperref[appendix:tables]{Appendix A} -- which contains information about all the other datasets analyzed in this work -- reveals a general trend toward higher or lower values of $\sigma_8$ when including or excluding information at \texttt{low-z}.

\subsection{Relativistic Degrees of Freedom}
The next step always involves modifications to the neutrino sector. In particular, we focus on the effective number of relativistic degrees of freedom at recombination, $N_{\rm eff}$. Within the Standard Model of particle physics, this parameter is predicted to be $N_{\rm eff} \simeq 3.04$. However, our analysis treats it as a free parameter in the cosmological model to be constrained by data.

In this case, the result obtained from the consensus dataset ($\tau = 0.080 \pm 0.012$) is identical to the one obtained within the standard cosmological model and so are the general conclusions. In contrast, the value obtained from \texttt{ACT(DR4+DR6)+low-z} shows a slight shift to larger values, accompanied by an increase in uncertainties: $\tau = 0.086 \pm 0.017$.

Regarding the value of $N_{\rm eff}$ obtained trough \texttt{TTTEEE+lensing+low-z}, we get $N_{\rm eff} = 2.95 \pm 0.16$ (see also \autoref{tab.results.lcdm_nnu} in \hyperref[appendix:tables]{Appendix A}) , in excellent agreement with the baseline value. On the other hand, as is well known, ACT favors a lower amount of radiation in the early universe, typically resulting in an anomaly in the value of $N_{\rm eff}$ at approximately $2.5\sigma$~\cite{ACT:2020gnv,Giare:2022rvg,Giare:2023xoc}. However, it is worth noting that the ACT-based constraint reads $N_{\rm eff} = 2.65 \pm 0.27$. Therefore, by constraining $\tau$ independently from Planck, the disagreement of ACT with the predictions of the standard model is reduced from  $2.5$ to $1.5$ standard deviations. This is due to a combined effect of increased uncertainties and a genuine shift in the central value of this parameter.

\subsection{Running of Inflationary perturbations}
We include the running of the spectral index, $\alpha_s = dn_s/d\log k$, in the cosmological model. In this case, the results for $\tau$ from \texttt{TTTEEE+lensing+low-z} measurements are $\tau = 0.079 \pm 0.012$, practically identical to $\Lambda$CDM. On the other hand, the constraints based on ACT measurements are $\tau = 0.055^{+0.018}_{-0.020}$. Despite the considerable uncertainties, the central value of the optical depth at reionization in this case appears to be closer to the \texttt{Planck-2108} results ($\tau=0.0553\pm 0.0077$). This shift to smaller values is attributed to the anti-correlation between $\tau$ and $\alpha_s$ depicted in \autoref{fig:tau.correlation}. The well-known preference of ACT for positive values of the latter~\cite{ACT:2020gnv,Forconi:2021que,Giare:2022rvg,Giare:2023xoc} pushes $\tau$ towards lower values.

This is also confirmed by looking at the results in \hyperref[appendix:tables]{Appendix A}, \autoref{tab.results.lcdm_nrun}. As one can notice, from the consensus dataset we measure $\alpha_s = 0.0018 \pm 0.0080$, in perfect agreement with $\alpha = 0$. Instead, from \texttt{ACT(DR4+DR6)+low-z}, we have $\alpha = 0.069 \pm 0.025$, confirming a general preference for $\alpha > 0$ at $\sim 2.8\sigma$.

\subsection{Curvature}
Allowing $\Omega_k$ to vary in the model, from the consensus dataset, once again, the constraints obtained for the optical depth at reionization are practically identical to those obtained in $\Lambda$CDM: $\tau = 0.079 \pm 0.013$. This confirms the robustness of this measurement when extending the cosmological model. Conversely, for \texttt{ACT(DR4+DR6)+low-z}, we obtain $\tau = 0.084 \pm 0.022$, i.e., significantly larger uncertainty margins compared to those obtained in the standard case.

Regarding the parameter $\Omega_k$, the inclusion of \texttt{low-z} data and precise lensing spectrum measurements impose strong constraints on this parameter and all combinations of data presented in \hyperref[appendix:tables]{Appendix A}, \autoref{tab.results.lcdm_omk} show excellent agreement with a flat Universe $\Omega_k=0$.

\subsection{Dynamical and non-Dynamical Dark Energy}
Now, we turn our attention to the study of extensions related to the DE sector of the theory. In this case, we examine two different extensions.

Firstly, we consider the $w$CDM model, where we generalize the cosmological constant $\Lambda$ to the case where DE is described by a non-dynamical (i.e., time-independent) equation of state $w$. In this case, from \texttt{TTTEEE+lensing+low-z}, we obtain a significantly higher value for the optical depth at reionization, $\tau = 0.090 \pm 0.014$, together with  a slight increase in uncertainty. A similar pattern is also observed by \texttt{ACT(DR4+DR6)+low-z}, yielding $\tau = 0.098 \pm 0.022$. Due to the broader error bars, these results do not exacerbate the difference with the results obtained from \texttt{Planck-2018} ($\tau=0.0524\pm 0.0074$). However, it is worth noting that this shift is due to the positive correlation between $\tau$ and $w$, as documented in some previous works~\cite{Escamilla:2023oce}. In particular, when the equation of state is allowed to vary, higher (lower) values of $\tau$ shift the results for $w$ in the direction of quintessential (phantom) DE. This is evident from the two-dimensional correlations shown in \autoref{fig:tau.correlation} and is corroborated by the results in \hyperref[appendix:tables]{Appendix A}, \autoref{tab.results.wcdm}. Specifically, both for the consensus dataset ($w = -0.967 \pm 0.027$) and for the constraints based on ACT ($w = -0.951 \pm 0.032$), we observe a shift towards $w > -1$ by $1.2\sigma$ and $1.53\sigma$, respectively.

Secondly, we consider a dynamic parametrization for the equation of state, meaning that the value of $w$ is allowed to vary with the scale factor $a$. In particular, we adopt the Chevallier-Polarski-Linder parametrization~\cite{Chevallier:2000qy,Linder:2002et}:
\begin{equation}
w(a) = w_0 + w_a (1-a)~,
\end{equation}
where $w_0$ is the present value $w(a=1)$, and $w_a$ is another free parameter such that $d w /\left.d \ln (1+z)\right|_{z=1}=w_a / 2$. Notice that the corresponding cosmological model is labeled as $w_0w_a$CDM. In this model (which has, therefore, 8 free parameters), we obtain $\tau = 0.074 \pm 0.017$ for \texttt{TTTEEE+lensing+low-z} and $\tau = 0.076^{+0.028}_{-0.036}$ for \texttt{ACT(DR4+DR6)+low-z}. Both datasets show a slight shift towards lower values of $\tau$, although the amplified error bars make this shift not statistically significant. Regarding the correlation between $\tau$, $w_0$, and $w_a$, \autoref{fig:tau.correlation} shows that in this case $\tau$ and $w_0$  becomes anti-correlated and increasing $\tau$ shifts the results towards $w_0<-1$. Conversely, $\tau$ and $w_a$ exhibit a strong positive correlation, and positive values of $w_a$ are only allowed if they result in large values of $\tau$. Looking at the results in  \hyperref[appendix:tables]{Appendix A}, \autoref{tab.results.w0wacdm}, it is also interesting to note that constraints on $w_0$ show a preference in favor of $w_0 > -1$ with a statistical significance ranging between $1.5$ and $2$ standard deviations. In particular, this holds true for \texttt{TTTEEE+lensing+low-z} (from which we obtain $w_0 = -0.880 \pm 0.062$) and for \texttt{ACT(DR4+DR6)+low-z} (from which we obtain $w_0 = -0.891 \pm 0.067$). Notably, the preference for $w_0 > -1$ is accompanied by an inclination towards a dynamic nature of the equation of state only in \texttt{TTTEEE+lensing+low-z} where it translates into a constraint $w_a = -0.41 \pm 0.27$ (i.e., $w_a \neq 0$ at $1.5\sigma$). Instead, for \texttt{ACT(DR4+DR6)+low-z}, $w_a = -0.36 \pm 0.34$ is consistent with the non-dynamical case essentially within one standard deviation.

\subsection{Lensing Amplitude}
We conclude the analysis of the extended models by revisiting the parameter $A_{\rm lens}$, whose physical implications have been extensively described in \autoref{sec:motivations}.

Our results once again confirm that, in absence of large-scale data, $A_{\rm lens}$ is the parameter with the most significant impact on $\tau$. In particular, the constraint on the optical depth to reionization obtained from the consensus dataset essentially becomes inconclusive: $\tau=0.099 \pm 0.034$, and the error bars become too wide to draw meaningful conclusions. The same argument applies to the result obtained for \texttt{ACT(DR4+DR6)+low-z}, which yields $\tau=0.083^{+0.033}_{-0.046}$ at the 68\% CL, while only an upper bound can be derived for the same parameter within the 95\% CL ($\tau<0.146$). However, all the datasets are in excellent agreement with $A_{\rm lens}=1$, see also \hyperref[appendix:tables]{Appendix A}, \autoref{tab.results.lcdm_Alens}

Overall, the analysis performed in this section demonstrates that, when extending the cosmological model, obtaining precise measurements of $\tau$ without \texttt{lowE} data may become a significantly more challenging endeavor than in the standard cosmological model. Looking on the bright side, for the consensus dataset, the result obtained within $\Lambda$CDM remains stable in the majority of the extended cosmologies. However, we observe a few exceptions when considering $A_{\rm lens}$, $\sum m_{\nu}$ and extensions related to DE. In these latter cases, the correlation between $\tau$ and the other parameters can lead to significant shifts as well as to a general loss of precision. Furthermore, it is worth highlighting that all observed shifts systematically tend towards higher values of $\tau$ than those obtained when including \texttt{lowE}. This frequently reflects in the constraints on other beyond-$\Lambda$CDM parameters. Hence, remaining agnostic about Planck polarization measurements at large angular scales, we can certainly conclude that an overall calibration of $\tau$ is crucial when studying extended cosmologies and, in general, theoretical models more complex than $\Lambda$CDM.

\section{Conclusions}
\label{sec:Conclusion}

The observational constraints on the optical depth at reionization $\tau$ have changed quite significantly over time due to a better understanding of foreground contamination and improved data accuracy. The state-of-the-art bounds derived by the Planck Collaboration represent a culmination, allowing us to determine this parameter with a relative precision of $\sim 14$\%.

However, it should be noted that the Planck constraints on $\tau$ are widely based on E-mode polarization measurements at $\ell\le 30$ (i.e., \texttt{lowE} data). Relying so much on one single dataset can be imprudent, especially considering the challenges in obtaining precise CMB measurements at large angular scales. As well known, low-multipole temperature data (i.e., \texttt{lowT} data) show unexpected features and correlations for multipoles $\ell < 10$, whereas the $C_{\ell}^{\mathrm{TE}}$ spectrum exhibits excess variance compared to simulations at $\ell\lesssim 20$, exactly the scales the signal should be dominated by the optical depth at reionization.

Similarly, when evaluating large-scale E-mode polarization measurements (i.e., the \texttt{lowE} data), it becomes crucial to assess their robustness as well as their consistency with other datasets. In \autoref{sec:motivations}, we highlight five independent reasons why we believe a degree of caution is advisable when dealing with \texttt{lowE} data. First and foremost, it is a fact that the amplitude of the spectrum of E-mode polarization at $\ell \le 30$ is extremely small, on the order of $10^{-3} - 10^{-2}$ $\mu K^2$, bringing the signal close to the experimental sensitivity. This makes any statistical fluctuation or lack of understanding of the foreground potentially crucial for the measurement of $\tau$. Additionally, when excluding \texttt{lowE} (and possibly \texttt{lowT}) data from the analysis, $\tau$ consistently shows correlations with anomalous parameters such as the lensing amplitude $A_{\mathrm{lens}}$, the curvature parameter $\Omega_k$, and the dark energy equation of state $w$. These correlations systematically suggest that slightly higher values of the optical depth $\tau\sim 0.07 - 0.08$ could significantly alleviate or even eliminate all anomalies encountered in the Planck data when extending the cosmological model (see also \autoref{fig:Alens} and \autoref{fig:omk_w}). Finally, when considering Planck-independent experiments probing small scales in the CMB, (Planck-based) priors for $\tau$ are usually assumed, as it is believed a priori that it is not possible to constrain this parameter without large-scale temperature and polarization data.

All these arguments emphasize the need to to obtain competitive measurements of the optical depth at reionization without relying on large-scale CMB data that could be used to cross-check the state-of-the-art results and dispel any doubts about using large-scale CMB measurements.

In this paper we have extensively studied this possibility by using different combinations of Planck and ACT temperature and polarization data at $\ell > 30$, Planck and ACT reconstructions of the lensing potential, Baryon Acoustic Oscillations measurements from BOSS and eBOSS surveys, and Type-Ia supernova data from the PantheonPlus sample. 

For the $\Lambda$CDM model, our results can be found in \autoref{tab.results.lcdm}, \autoref{fig:tau.lcdm} and \autoref{fig:tau.lcdm_Tplot}. The most relevant conclusions are as follows:

\begin{itemize}

\item Considering CMB measurements at $\ell>30$ and local Universe probes, we can achieve constraints on $\tau$ that are independent from large-scale temperature and polarization measurements. From \texttt{TTTEEE+lensing+low-z} (identified as the consensus dataset), we obtain the most constraining result $\tau = 0.080 \pm 0.012$. This result conclusively demonstrates that not only is it possible to measure $\tau$ without large-scale CMB data, but also that the relative precision remains comparable to the constraints based on large-scale E-mode polarization ($\tau = 0.054 \pm 0.008$).

\item The results on $\tau$ derived considering only Planck data at $\ell > 30$ show a trend towards slightly larger values of $\tau$ compared to those derived including measurements at $\ell \le 30$. In particular, for the consensus dataset, we report a $1.8\sigma$ shift towards larger values of $\tau \sim 0.08$. This tiny shift is confirmed by different Planck likelihoods that employ diverse techniques for handling foregrounds and reducing noise at small scales, such as \texttt{plik}, \texttt{Camspec}, and \texttt{HiLLiPoP}, see also \hyperref[sec.AppendixB]{Appendix B}.

\item The shift towards significantly larger $\tau$ can have mild to moderate implications for other cosmological parameters and tensions. On the one hand, as we argued in \autoref{sec:motivations}, larger values $\tau \sim 0.07, 0.08$ can help alleviate anomalies observed in the Planck data, most notably the lensing and curvature anomalies. On the other hand, in \autoref{sec:LCDM}, we point out that the $1.8\sigma$ shift in the value of $\tau$ observed excluding large-scale CMB data is responsible for a $1.8\sigma$ shift towards higher values of the amplitude of primordial inflationary perturbations $A_s$ compared to \texttt{Planck-2018}, and consequently to higher values of the parameter $\sigma_8$, possibly exacerbating the difference between CMB and WL surveys.

\item Notably, using only ACT-based temperature, polarization, and lensing data in combination with local universe measurements, from the dataset \texttt{ACT(DR4+DR6)+low-z}, we can derive an independent measurement $\tau = 0.076 \pm 0.015$, which does not rely on Planck. The relative precision of this result is competitive with the value extracted from Planck (both with and without considering large-scale polarization). Therefore, not only is it possible to constrain $\tau$ with small-scale CMB data, but also it is possible to obtain independent measurements from ACT without a need to consider Planck-based priors on this parameter, as commonly done in the literature.

\item Planck-independent measurement of $\tau$ based on small-scale data confirms the overall preference towards slightly larger values, lending weight to the robustness of the determination of this parameter without large-scale data.

\end{itemize}


Finally, in \autoref{sec:extensions}, we have tested how the results change by extending the cosmological model. Considering the usual array of extended cosmologies typically analyzed when deriving constraints beyond $\Lambda$CDM, we find that:

\begin{itemize}

\item Considering the effective number of relativistic degrees of freedom in the early Universe $N_{\mathrm{eff}}$, the spectral index running $\alpha_s$, and the curvature density parameter $\Omega_k$ as free parameters, the constraints on $\tau$ remain largely consistent with those inferred within the standard cosmological model. The inclusion of these additional parameters does not alter the results derived within $\Lambda$CDM from small-scale CMB data (both alone and in conjunction with low-redshift data), nor does it compromise the precision of measurements obtained.

\item In contrast, obtaining precise measurements of $\tau$ may become significantly more challenging when considering the total neutrino mass $\sum m_{\nu}$ or the lensing amplitude $A_{\rm lens}$ as free parameters, as well as in extensions involving modifications to the dark energy (DE) sector of the theory. 

\end{itemize}

We refer to \autoref{tab.results.extensions} and \autoref{fig:tau.extensions} for a summary of the results.

\begin{acknowledgments}
EDV is supported by a Royal Society Dorothy Hodgkin Research Fellowship. AM is supported by "Theoretical Astroparticle Physics" (TAsP), iniziativa specifica INFN. The work of A.M. is partially supported by the research grant number 2022E2J4RK ``PANTHEON:
Perspectives in Astroparticle and Neutrino THEory with Old and New messengers''
under the program PRIN 2022 funded by the Italian Ministero dell’Universit\`a e della Ricerca (MUR).
This article is based upon work from COST Action CA21136 Addressing observational tensions in cosmology with systematics and fundamental physics (CosmoVerse) supported by COST (European Cooperation in Science and Technology). We acknowledge IT Services at The University of Sheffield for the provision of services for High Performance Computing.
\end{acknowledgments}

\begin{widetext}
\appendix
\section{Tables with full results}
\label{appendix:tables}

\begin{table*}[htbp!]
\begin{center}
\renewcommand{\arraystretch}{1.5}
\resizebox{\textwidth}{!}{
\begin{tabular}{l c c c c c c c c c c c c c c c }
\hline
\textbf{Parameter} & \textbf{ 	\texttt{TT+lensing} }  & \textbf{ 	\texttt{TT+lensing+low-z} } & \textbf{ 	\texttt{TTTEEE+lensing} } & \textbf{ 	\texttt{TTTEEE+lensing+low-z} } & \textbf{ \texttt{ACT(DR4+DR6)} } & \textbf{ \texttt{ACT(DR4+DR6)+low-z}} \\ 
\hline\hline

 $ \tau_\mathrm{reio}  $ & $  0.088\pm 0.025\, ( 0.088^{+0.050}_{-0.049} ) $ & $  0.100\pm 0.019\, ( 0.100^{+0.042}_{-0.040} ) $ & $  0.092\pm 0.018\, ( 0.092^{+0.034}_{-0.035} ) $ & $  0.095\pm 0.016\, ( 0.095^{+0.034}_{-0.033} ) $ & $  0.113^{+0.041}_{-0.047}\, ( 0.113^{+0.082}_{-0.086} ) $ & $  0.133\pm 0.034\, ( 0.133^{+0.072}_{-0.071} ) $ \\ 
$ \sum m_\nu \, [eV]  $ & $ < 0.470 $ & $ < 0.275 $ & $ < 0.441 $ & $ < 0.228 $ & $ < 0.863 $ & $ < 0.476 $ \\ 
$ \Omega_\mathrm{b} h^2  $ & $  0.02220\pm 0.00028 $ & $  0.02239\pm 0.00021 $ & $  0.02244\pm 0.00017 $ & $  0.02254\pm 0.00014 $ & $  0.02150\pm 0.00031 $ & $  0.02158\pm 0.00030 $ \\ 
$ \Omega_\mathrm{c} h^2  $ & $  0.1192\pm 0.0027 $ & $  0.1171\pm 0.0014 $ & $  0.1189\pm 0.0016 $ & $  0.1178\pm 0.0011 $ & $  0.1191\pm 0.0047 $ & $  0.1153\pm 0.0023 $ \\ 
$ 100\theta_\mathrm{MC}  $ & $  1.04086\pm 0.00054 $ & $  1.04117\pm 0.00043 $ & $  1.04096\pm 0.00033 $ & $  1.04112\pm 0.00029 $ & $  1.04193\pm 0.00077 $ & $  1.04239\pm 0.00063 $ \\ 
$ n_\mathrm{s}  $ & $  0.9654\pm 0.0083 $ & $  0.9717\pm 0.0051 $ & $  0.9674\pm 0.0054 $ & $  0.9709\pm 0.0043 $ & $  1.001\pm 0.016 $ & $  1.011\pm 0.014 $ \\ 
$ \log(10^{10} A_\mathrm{s})  $ & $  3.108\pm 0.044 $ & $  3.127\pm 0.035 $ & $  3.115\pm 0.032 $ & $  3.119\pm 0.028 $ & $  3.149\pm 0.068 $ & $  3.180\pm 0.058 $ \\ 
\hline
$ H_0  $ & $  65.9\pm 1.7 $ & $  67.66\pm 0.54 $ & $  66.4\pm 1.2 $ & $  67.69\pm 0.49 $ & $  63.9\pm 2.6 $ & $  67.05\pm 0.57 $ \\ 
$ \sigma_8  $ & $  0.793^{+0.026}_{-0.024} $ & $  0.815\pm 0.010 $ & $  0.798^{+0.023}_{-0.020} $ & $  0.8179\pm 0.0095 $ & $  0.775\pm 0.035 $ & $  0.819\pm 0.015 $ \\ 

\hline \hline
\end{tabular} }
\end{center}
\caption{Results at 68\% (95\%) CL on cosmological parameters obtained within $\Lambda$CDM+$\sum m_{\nu}$. }
\label{tab.results.lcdm_mnu}
\end{table*}
\begin{table*}[htbp!]
\begin{center}
\renewcommand{\arraystretch}{1.5}
\resizebox{\textwidth}{!}{
\begin{tabular}{l c c c c c c c c c c c c c c c }
\hline
\textbf{Parameter} & \textbf{ 	\texttt{TT+lensing} }  & \textbf{ 	\texttt{TT+lensing+low-z} } & \textbf{ 	\texttt{TTTEEE+lensing} } & \textbf{ 	\texttt{TTTEEE+lensing+low-z} } & \textbf{ \texttt{ACT(DR4+DR6)} } & \textbf{ \texttt{ACT(DR4+DR6)+low-z}} \\
\hline\hline

$ \tau_\mathrm{reio}  $ & $  0.072^{+0.022}_{-0.053}\, (< 0.136 ) $ & $  0.078\pm 0.013\, ( 0.078^{+0.025}_{-0.025} ) $ & $  0.073\pm 0.017\, ( 0.073^{+0.034}_{-0.033} ) $ & $  0.080\pm 0.012\, ( 0.080^{+0.023}_{-0.023} ) $ & $ < 0.0624\, (< 0.117 ) $ & $  0.086\pm 0.017\, ( 0.086^{+0.034}_{-0.033} ) $ \\ 
$ N_\mathrm{eff}  $ & $  2.93\pm 0.39 $ & $  3.03\pm 0.21 $ & $  2.87\pm 0.20 $ & $  2.95\pm 0.16 $ & $  2.26\pm 0.38 $ & $  2.65\pm 0.27 $ \\ 
$ \Omega_\mathrm{b} h^2  $ & $  0.02218\pm 0.00055 $ & $  0.02229\pm 0.00022 $ & $  0.02231\pm 0.00026 $ & $  0.02243\pm 0.00018 $ & $  0.02091\pm 0.00046 $ & $  0.02129\pm 0.00037 $ \\ 
$ \Omega_\mathrm{c} h^2  $ & $  0.1173\pm 0.0037 $ & $  0.1182\pm 0.0036 $ & $  0.1165\pm 0.0029 $ & $  0.1170\pm 0.0028 $ & $  0.1094\pm 0.0051 $ & $  0.1118\pm 0.0049 $ \\ 
$ 100\theta_\mathrm{MC}  $ & $  1.04117\pm 0.00057 $ & $  1.04107\pm 0.00058 $ & $  1.04130\pm 0.00044 $ & $  1.04125\pm 0.00043 $ & $  1.04320\pm 0.00089 $ & $  1.04288\pm 0.00083 $ \\ 
$ n_\mathrm{s}  $ & $  0.962\pm 0.024 $ & $  0.9673\pm 0.0077 $ & $  0.960\pm 0.010 $ & $  0.9654\pm 0.0067 $ & $  0.949\pm 0.030 $ & $  0.981\pm 0.018 $ \\ 
$ \log(10^{10} A_\mathrm{s})  $ & $  3.071^{+0.076}_{-0.086} $ & $  3.086\pm 0.022 $ & $  3.071\pm 0.034 $ & $  3.088\pm 0.021 $ & $  3.006^{+0.063}_{-0.079} $ & $  3.082\pm 0.026 $ \\ 
\hline
$ H_0  $ & $  67.0^{+4.1}_{-4.7} $ & $  67.8\pm 1.3 $ & $  66.6\pm 1.7 $ & $  67.4\pm 1.1 $ & $  61.5^{+3.5}_{-4.0} $ & $  65.4\pm 1.5 $ \\ 
$ \sigma_8  $ & $  0.818^{+0.029}_{-0.033} $ & $  0.824\pm 0.010 $ & $  0.816\pm 0.014 $ & $  0.8222\pm 0.0095 $ & $  0.792^{+0.026}_{-0.031} $ & $  0.823\pm 0.014 $ \\

\hline \hline
\end{tabular} }
\end{center}
\caption{Results at 68\% (95\%) CL on cosmological parameters obtained within $\Lambda$CDM+$N_{\rm{eff}}$. }
\label{tab.results.lcdm_nnu}
\end{table*}
\begin{table*}[htbp!]
\begin{center}
\renewcommand{\arraystretch}{1.5}
\resizebox{\textwidth}{!}{
\begin{tabular}{l c c c c c c c c c c c c c c c }
\hline
\textbf{Parameter} & \textbf{ 	\texttt{TT+lensing} }  & \textbf{ 	\texttt{TT+lensing+low-z} } & \textbf{ 	\texttt{TTTEEE+lensing} } & \textbf{ 	\texttt{TTTEEE+lensing+low-z} } & \textbf{ \texttt{ACT(DR4+DR6)} } & \textbf{ \texttt{ACT(DR4+DR6)+low-z}} \\ 
\hline\hline

$ \tau_\mathrm{reio}  $ & $  0.072^{+0.026}_{-0.031}\, ( 0.072^{+0.051}_{-0.055} ) $ & $  0.077\pm 0.013\, ( 0.077^{+0.025}_{-0.025} ) $ & $  0.077\pm 0.016\, ( 0.077^{+0.032}_{-0.032} ) $ & $  0.079\pm 0.012\, ( 0.079^{+0.024}_{-0.023} ) $ & $ < 0.0569\, (< 0.103 ) $ & $  0.055^{+0.018}_{-0.020}\, ( 0.055^{+0.035}_{-0.037} ) $ \\ 
$ \alpha_s  $ & $  0.006\pm 0.010 $ & $  0.0059\pm 0.0094 $ & $  0.0019\pm 0.0080 $ & $  0.0018\pm 0.0080 $ & $  0.071\pm 0.025 $ & $  0.069\pm 0.025 $ \\ 
$ \Omega_\mathrm{b} h^2  $ & $  0.02218\pm 0.00033 $ & $  0.02222\pm 0.00023 $ & $  0.02247\pm 0.00018 $ & $  0.02249\pm 0.00015 $ & $  0.02131\pm 0.00032 $ & $  0.02133\pm 0.00031 $ \\ 
$ \Omega_\mathrm{c} h^2  $ & $  0.1189\pm 0.0027 $ & $  0.1184\pm 0.0011 $ & $  0.1187\pm 0.0016 $ & $  0.11855\pm 0.00099 $ & $  0.1201^{+0.0039}_{-0.0033} $ & $  0.1191\pm 0.0013 $ \\ 
$ 100\theta_\mathrm{MC}  $ & $  1.04094\pm 0.00054 $ & $  1.04100\pm 0.00042 $ & $  1.04104\pm 0.00033 $ & $  1.04106\pm 0.00029 $ & $  1.04217\pm 0.00070 $ & $  1.04227\pm 0.00063 $ \\ 
$ n_\mathrm{s}  $ & $  0.9671\pm 0.0077 $ & $  0.9683\pm 0.0043 $ & $  0.9684\pm 0.0054 $ & $  0.9690\pm 0.0042 $ & $  0.968\pm 0.018 $ & $  0.971\pm 0.016 $ \\ 
$ \log(10^{10} A_\mathrm{s})  $ & $  3.074\pm 0.046 $ & $  3.082\pm 0.024 $ & $  3.085\pm 0.029 $ & $  3.090\pm 0.022 $ & $  3.025^{+0.047}_{-0.060} $ & $  3.037\pm 0.030 $ \\ 
\hline
$ H_0  $ & $  67.6\pm 1.3 $ & $  67.80\pm 0.53 $ & $  67.92\pm 0.73 $ & $  68.00\pm 0.45 $ & $  66.9^{+1.3}_{-1.6} $ & $  67.25\pm 0.52 $ \\ 
$ \sigma_8  $ & $  0.822\pm 0.011 $ & $  0.8243\pm 0.0076 $ & $  0.8246\pm 0.0086 $ & $  0.8260\pm 0.0074 $ & $  0.824^{+0.013}_{-0.016} $ & $  0.827\pm 0.011 $ \\ 

\hline \hline
\end{tabular} }
\end{center}
\caption{ Results at 68\% (95\%) CL on cosmological parameters obtained within $\Lambda$CDM+$\alpha_s$.  }
\label{tab.results.lcdm_nrun}
\end{table*}
\begin{table*}[htbp!]
\begin{center}
\renewcommand{\arraystretch}{1.5}
\resizebox{\textwidth}{!}{
\begin{tabular}{l c c c c c c c c c c c c c c c }
\hline
\textbf{Parameter} & \textbf{ 	\texttt{TT+lensing} }  & \textbf{ 	\texttt{TT+lensing+low-z} } & \textbf{ 	\texttt{TTTEEE+lensing} } & \textbf{ 	\texttt{TTTEEE+lensing+low-z} } & \textbf{ \texttt{ACT(DR4+DR6)} } & \textbf{ \texttt{ACT(DR4+DR6)+low-z}} \\ 
\hline\hline

$ \tau_\mathrm{reio}  $ & $  0.093^{+0.032}_{-0.070}\, (< 0.172 ) $ & $  0.078\pm 0.016\, ( 0.078^{+0.031}_{-0.031} ) $ & $  0.093^{+0.039}_{-0.055}\, (< 0.163 ) $ & $  0.079\pm 0.013\, ( 0.079^{+0.025}_{-0.024} ) $ & $ < 0.151\, (< 0.252 ) $ & $  0.084\pm 0.022\, ( 0.084^{+0.045}_{-0.043} ) $ \\ 
$ \Omega_k  $ & $  0.003^{+0.014}_{-0.012} $ & $  0.0002\pm 0.0025 $ & $  0.003^{+0.014}_{-0.012} $ & $  0.0004\pm 0.0018 $ & $  0.002^{+0.025}_{-0.021} $ & $  -0.0015\pm 0.0036 $ \\ 
$ \Omega_\mathrm{b} h^2  $ & $  0.02232\pm 0.00026 $ & $  0.02230\pm 0.00025 $ & $  0.02249\pm 0.00016 $ & $  0.02249\pm 0.00016 $ & $  0.02163\pm 0.00031 $ & $  0.02162\pm 0.00029 $ \\ 
$ \Omega_\mathrm{c} h^2  $ & $  0.1183\pm 0.0025 $ & $  0.1185\pm 0.0023 $ & $  0.1187\pm 0.0016 $ & $  0.1188\pm 0.0015 $ & $  0.1166\pm 0.0046 $ & $  0.1174\pm 0.0036 $ \\ 
$ 100\theta_\mathrm{MC}  $ & $  1.04102\pm 0.00052 $ & $  1.04100\pm 0.00050 $ & $  1.04104\pm 0.00033 $ & $  1.04103\pm 0.00031 $ & $  1.04229\pm 0.00075 $ & $  1.04226\pm 0.00071 $ \\ 
$ n_\mathrm{s}  $ & $  0.9687\pm 0.0078 $ & $  0.9677\pm 0.0069 $ & $  0.9690\pm 0.0053 $ & $  0.9682\pm 0.0050 $ & $  1.006\pm 0.017 $ & $  1.003\pm 0.015 $ \\ 
$ \log(10^{10} A_\mathrm{s})  $ & $  3.116\pm 0.091 $ & $  3.086\pm 0.026 $ & $  3.118\pm 0.083 $ & $  3.090\pm 0.022 $ & $  3.16^{+0.15}_{-0.17} $ & $  3.092\pm 0.035 $ \\ 
\hline
$ H_0  $ & $  70.4\pm 6.6 $ & $  67.94\pm 0.59 $ & $  70.5\pm 6.2 $ & $  68.09\pm 0.59 $ & $  72\pm 10 $ & $  67.38\pm 0.63 $ \\ 
$ \sigma_8  $ & $  0.838\pm 0.043 $ & $  0.8247\pm 0.0077 $ & $  0.840\pm 0.040 $ & $  0.8265\pm 0.0077 $ & $  0.867\pm 0.075 $ & $  0.837\pm 0.010 $ \\

\hline \hline
\end{tabular} }
\end{center}
\caption{ Results at 68\% (95\%) CL on cosmological parameters obtained within $\Lambda$CDM+$\Omega_k$. }
\label{tab.results.lcdm_omk}
\end{table*}

\begin{table*}[htbp!]
\begin{center}
\renewcommand{\arraystretch}{1.5}
\resizebox{\textwidth}{!}{
\begin{tabular}{l c c c c c c c c c c c c c c c }
\hline
\textbf{Parameter} & \textbf{ 	\texttt{TT+lensing} }  & \textbf{ 	\texttt{TT+lensing+low-z} } & \textbf{ 	\texttt{TTTEEE+lensing} } & \textbf{ 	\texttt{TTTEEE+lensing+low-z} } & \textbf{ \texttt{ACT(DR4+DR6)} } & \textbf{ \texttt{ACT(DR4+DR6)+low-z}} \\ 
\hline\hline

 $ \tau_\mathrm{reio}  $ & $  0.072^{+0.026}_{-0.040}\, (< 0.128 ) $ & $  0.095\pm 0.018\, ( 0.095^{+0.035}_{-0.034} ) $ & $  0.073^{+0.023}_{-0.029}\, ( 0.073^{+0.052}_{-0.050} ) $ & $  0.090\pm 0.014\, ( 0.090^{+0.028}_{-0.028} ) $ & $  0.084^{+0.029}_{-0.059}\, (< 0.160 ) $ & $  0.098\pm 0.022\, ( 0.098^{+0.043}_{-0.041} ) $ \\ 
$ w  $ & $  -1.27\pm 0.40 $ & $  -0.957\pm 0.030 $ & $  -1.27\pm 0.40 $ & $  -0.967\pm 0.027 $ & $  -1.33^{+0.42}_{-0.48} $ & $  -0.951\pm 0.032 $ \\ 
$ \Omega_\mathrm{b} h^2  $ & $  0.02228\pm 0.00027 $ & $  0.02244\pm 0.00022 $ & $  0.02248\pm 0.00017 $ & $  0.02257\pm 0.00015 $ & $  0.02162\pm 0.00030 $ & $  0.02165\pm 0.00030 $ \\ 
$ \Omega_\mathrm{c} h^2  $ & $  0.1186\pm 0.0026 $ & $  0.1168\pm 0.0016 $ & $  0.1188\pm 0.0016 $ & $  0.1176\pm 0.0012 $ & $  0.1173\pm 0.0041 $ & $  0.1166\pm 0.0019 $ \\ 
$ 100\theta_\mathrm{MC}  $ & $  1.04100\pm 0.00052 $ & $  1.04123\pm 0.00044 $ & $  1.04103\pm 0.00032 $ & $  1.04116\pm 0.00030 $ & $  1.04226\pm 0.00075 $ & $  1.04230\pm 0.00062 $ \\ 
$ n_\mathrm{s}  $ & $  0.9674\pm 0.0079 $ & $  0.9726\pm 0.0056 $ & $  0.9679\pm 0.0053 $ & $  0.9714\pm 0.0046 $ & $  1.004\pm 0.015 $ & $  1.005\pm 0.013 $ \\ 
$ \log(10^{10} A_\mathrm{s})  $ & $  3.075^{+0.056}_{-0.065} $ & $  3.117\pm 0.031 $ & $  3.077^{+0.045}_{-0.053} $ & $  3.109\pm 0.026 $ & $  3.092^{+0.073}_{-0.085} $ & $  3.118\pm 0.037 $ \\ 
\hline
$ H_0  $ & $  77^{+20}_{-8} $ & $  67.30\pm 0.65 $ & $  77^{+20}_{-8} $ & $  67.40\pm 0.66 $ & $ > 72.9 $ & $  66.87\pm 0.67 $ \\ 
$ \sigma_8  $ & $  0.89\pm 0.10 $ & $  0.8191\pm 0.0084 $ & $  0.89\pm 0.10 $ & $  0.8210\pm 0.0084 $ & $  0.93^{+0.13}_{-0.11} $ & $  0.832\pm 0.011 $ \\ 

\hline \hline
\end{tabular} }
\end{center}
\caption{ Results at 68\% (95\%) CL on cosmological parameters obtained within $w$CDM.  }
\label{tab.results.wcdm}
\end{table*}
\begin{table*}[htbp!]
\begin{center}
\renewcommand{\arraystretch}{1.5}
\resizebox{\textwidth}{!}{
\begin{tabular}{l c c c c c c c c c c c c c c c }
\hline
\textbf{Parameter} & \textbf{ 	\texttt{TT+lensing} }  & \textbf{ 	\texttt{TT+lensing+low-z} } & \textbf{ 	\texttt{TTTEEE+lensing} } & \textbf{ 	\texttt{TTTEEE+lensing+low-z} } & \textbf{ \texttt{ACT(DR4+DR6)} } & \textbf{ \texttt{ACT(DR4+DR6)+low-z}} \\  
\hline\hline

$ \tau_\mathrm{reio}  $ & $  0.069^{+0.025}_{-0.040}\, (< 0.123 ) $ & $  0.073^{+0.023}_{-0.027}\, ( 0.073^{+0.048}_{-0.049} ) $ & $  0.070^{+0.022}_{-0.027}\, ( 0.070^{+0.049}_{-0.047} ) $ & $  0.074\pm 0.017\, ( 0.074^{+0.034}_{-0.033} ) $ & $  0.082^{+0.026}_{-0.060}\, (< 0.156 ) $ & $  0.076^{+0.028}_{-0.036}\, ( 0.076^{+0.057}_{-0.062} ) $ \\ 
$ w_0  $ & $  -0.95\pm 0.67 $ & $  -0.881\pm 0.065 $ & $  -0.97\pm 0.66 $ & $  -0.880\pm 0.062 $ & $  -1.02\pm 0.64 $ & $  -0.891\pm 0.067 $ \\ 
$ w_{a}  $ & $ < -0.334 $ & $  -0.41\pm 0.32 $ & $ < -0.336 $ & $  -0.41\pm 0.27 $ & $ < -0.421 $ & $  -0.36\pm 0.34 $ \\ 
$ \Omega_\mathrm{b} h^2  $ & $  0.02227\pm 0.00027 $ & $  0.02229\pm 0.00024 $ & $  0.02248\pm 0.00017 $ & $  0.02248\pm 0.00016 $ & $  0.02162\pm 0.00029 $ & $  0.02162\pm 0.00029 $ \\ 
$ \Omega_\mathrm{c} h^2  $ & $  0.1187\pm 0.0026 $ & $  0.1186\pm 0.0021 $ & $  0.1188\pm 0.0016 $ & $  0.1188\pm 0.0014 $ & $  0.1173\pm 0.0041 $ & $  0.1187\pm 0.0027 $ \\ 
$ 100\theta_\mathrm{MC}  $ & $  1.04097\pm 0.00052 $ & $  1.04099\pm 0.00047 $ & $  1.04103\pm 0.00032 $ & $  1.04103\pm 0.00031 $ & $  1.04227\pm 0.00073 $ & $  1.04208\pm 0.00065 $ \\ 
$ n_\mathrm{s}  $ & $  0.9671\pm 0.0078 $ & $  0.9673\pm 0.0066 $ & $  0.9680\pm 0.0053 $ & $  0.9680\pm 0.0049 $ & $  1.004\pm 0.016 $ & $  0.9998\pm 0.014 $ \\ 
$ \log(10^{10} A_\mathrm{s})  $ & $  3.069^{+0.055}_{-0.064} $ & $  3.078\pm 0.041 $ & $  3.072^{+0.042}_{-0.049} $ & $  3.081\pm 0.031 $ & $  3.088^{+0.072}_{-0.085} $ & $  3.080\pm 0.050 $ \\ 
\hline
$ H_0  $ & $  75^{+20}_{-8} $ & $  67.38\pm 0.68 $ & $  75^{+20}_{-6} $ & $  67.50\pm 0.66 $ & $ > 70.4 $ & $  67.02\pm 0.68 $ \\ 
$ \sigma_8  $ & $  0.87\pm 0.12 $ & $  0.8197\pm 0.0085 $ & $  0.88\pm 0.12 $ & $  0.8211\pm 0.0083 $ & $  0.92^{+0.15}_{-0.13} $ & $  0.833\pm 0.011 $ \\  

\hline \hline
\end{tabular} }
\end{center}
\caption{Results at 68\% (95\%) CL on cosmological parameters obtained within $w_0w_a$CDM. }
\label{tab.results.w0wacdm}
\end{table*}
\begin{table*}[htbp!]
\begin{center}
\renewcommand{\arraystretch}{1.5}
\resizebox{\textwidth}{!}{
\begin{tabular}{l c c c c c c c c c c c c c c c }
\hline
\textbf{Parameter} & \textbf{ 	\texttt{TT+lensing} }  & \textbf{ 	\texttt{TT+lensing+low-z} } & \textbf{ 	\texttt{TTTEEE+lensing} } & \textbf{ 	\texttt{TTTEEE+lensing+low-z} } & \textbf{ \texttt{ACT(DR4+DR6)} } & \textbf{ \texttt{ACT(DR4+DR6)+low-z}} \\ 
\hline\hline

$ \tau_\mathrm{reio}  $ & $ < 0.116\, (< 0.177 ) $ & $  0.099\pm 0.034\, ( 0.099^{+0.065}_{-0.070} ) $ & $ < 0.113\, (< 0.174 ) $ & $  0.099\pm 0.034\, ( 0.099^{+0.064}_{-0.069} ) $ & $ < 0.150\, (< 0.327 ) $ & $  0.083^{+0.033}_{-0.046}\, (< 0.146 ) $ \\ 
$ A_{\rm lens}  $ & $  0.985\pm 0.099 $ & $  0.958\pm 0.069 $ & $  0.982\pm 0.096 $ & $  0.959\pm 0.068 $ & $  0.98^{+0.22}_{-0.19} $ & $  0.989\pm 0.078 $ \\ 
$ \Omega_\mathrm{b} h^2  $ & $  0.02232\pm 0.00027 $ & $  0.02228\pm 0.00020 $ & $  0.02249\pm 0.00017 $ & $  0.02248\pm 0.00014 $ & $  0.02164\pm 0.00030 $ & $  0.02161\pm 0.00029 $ \\ 
$ \Omega_\mathrm{c} h^2  $ & $  0.1181\pm 0.0026 $ & $  0.1185\pm 0.0012 $ & $  0.1186\pm 0.0016 $ & $  0.11869\pm 0.00099 $ & $  0.1157\pm 0.0044 $ & $  0.1188\pm 0.0014 $ \\ 
$ 100\theta_\mathrm{MC}  $ & $  1.04105\pm 0.00053 $ & $  1.04100\pm 0.00042 $ & $  1.04105\pm 0.00033 $ & $  1.04104\pm 0.00029 $ & $  1.04240\pm 0.00074 $ & $  1.04210\pm 0.00063 $ \\ 
$ n_\mathrm{s}  $ & $  0.9692\pm 0.0078 $ & $  0.9681\pm 0.0045 $ & $  0.9690\pm 0.0054 $ & $  0.9688\pm 0.0041 $ & $  1.007\pm 0.016 $ & $  0.999\pm 0.013 $ \\ 
$ \log(10^{10} A_\mathrm{s})  $ & $  3.110\pm 0.099 $ & $  3.128\pm 0.068 $ & $  3.111\pm 0.096 $ & $  3.130\pm 0.067 $ & $  3.17^{+0.18}_{-0.26} $ & $  3.095\pm 0.072 $ \\ 
\hline
$ H_0  $ & $  68.0\pm 1.2 $ & $  67.84\pm 0.53 $ & $  67.98\pm 0.72 $ & $  67.94\pm 0.45 $ & $  68.8\pm 1.8 $ & $  67.54\pm 0.52 $ \\ 
$ \sigma_8  $ & $  0.834\pm 0.041 $ & $  0.842\pm 0.029 $ & $  0.835^{+0.042}_{-0.047} $ & $  0.844\pm 0.028 $ & $  0.870^{+0.080}_{-0.11} $ & $  0.844\pm 0.030 $ \\

\hline \hline
\end{tabular} }
\end{center}
\caption{ Results at 68\% (95\%) CL on cosmological parameters obtained within $\Lambda$CDM+$A_{\rm lens}$.  }
\label{tab.results.lcdm_Alens}
\end{table*}

\clearpage

\section{Results from alternative high-$\ell$ Planck likelihoods}
\label{sec.AppendixB}
In this paper, we have examined the possibility of measuring the optical depth at reionization, $\tau$, using small-scale CMB data, either alone or in conjunction with low-redshift probes. We focused on two independent CMB experiments, Planck and ACT. For Planck, our results are based exclusively on the analysis of temperature and polarization spectra from the \texttt{plik} likelihood. We chose \texttt{plik} partly because it is the baseline likelihood used in the Planck-2018 papers and partly because it is widely adopted in the literature, enabling a direct comparison with other studies. However, over the years, the Planck data have been the subject of reanalyses. Specifically, as mentioned in \autoref{sec:Method}, the recent Planck PR4 \texttt{NPIPE} CMB map~\cite{Carron:2022eyg} incorporates several improvements, accounting for more sky area at high frequencies and addressing several enhancements in the processing of time-ordered data. This has led to a significant reduction in small-scale noise compared to \texttt{plik}. Following these overall improvements, novel likelihoods for temperature and polarization anisotropy spectra have been released~\cite{Rosenberg:2022sdy,Tristram:2023haj}, enhancing constraints on cosmological parameters, sometimes up to 10\%. Furthermore, these alternative likelihoods often employ different methods for foreground removal. All these aspects, particularly the reduced noise on a small scale and advancements in foreground modeling, can have significant implications for the constraints on $\tau$.

In this appendix, assuming a standard $\Lambda$CDM model of cosmology, we study what kind of constraints we can derive from small-scale data using alternative Planck likelihoods extracted from the \texttt{NPIPE} maps. Specifically, we consider the following two likelihoods:
\begin{itemize}
\item \texttt{CamSpec}~\cite{Rosenberg:2022sdy}, based on high-$\ell$ CMB temperature and polarization power spectra derived from the Planck PR4 \texttt{NPIPE} maps. We will refer to \texttt{CamSpec-TT} when we include only high-$\ell$ temperature data and \texttt{CamSpec-TTTEEE} when we use both high-$\ell$ temperature and polarization data.

\item \texttt{HiLLiPoP}~\cite{Tristram:2023haj}, always based on high-$\ell$ CMB temperature and polarization power spectra derived from the Planck PR4 \texttt{NPIPE} maps, but it relies on physical modeling of the foreground residuals in the spectral domain. We will refer to \texttt{HiLLiPoP-TT} when we use only high-$\ell$ temperature data and \texttt{HiLLiPoP-TTTEEE} when including high-$\ell$ polarization measurements, too.
\end{itemize}
Notice that these two likelihoods will always be used in combination with the \texttt{lensing} likelihood as well as \texttt{low-z} data, both detailed in \autoref{sec:Method}.

The results for all the 6 cosmological parameters based on the \texttt{CamSpec} likelihood are summarized in~\autoref{tab.results.lcdm.camspec}. Instead, the results obtained from \texttt{HiLLiPoP} are given in~\autoref{tab.results.lcdm.hillipop}.

\begin{table*}[!ht]
\begin{center}
\renewcommand{\arraystretch}{1.5}
\resizebox{\textwidth}{!}{
\begin{tabular}{l c c c c c c c c c c c c c c c }
\hline
\textbf{Parameter} & \textbf{ \texttt{Camspec-TT+lensing} } & \textbf{\texttt{Camspec-TT+lensing+low-z} } & \textbf{ \texttt{Camspec-TTTEEE+lensing} } & \textbf{ \texttt{Camspec-TTTEEE+lensing+low-z} } \\ 
\hline\hline

$ \tau_\mathrm{reio}  $ & $  0.083\pm 0.025\, ( 0.083^{+0.049}_{-0.048} ) $ & $  0.079\pm 0.012\, ( 0.079^{+0.024}_{-0.024} ) $ & $  0.069\pm 0.015\, ( 0.069^{+0.029}_{-0.029} ) $ & $  0.075\pm 0.011\, ( 0.075^{+0.023}_{-0.022} ) $ \\ 
$ \Omega_\mathrm{b} h^2  $ & $  0.02230\pm 0.00027 $ & $  0.02225\pm 0.00019 $ & $  0.02224\pm 0.00015 $ & $  0.02229\pm 0.00013 $ \\ 
$ \Omega_\mathrm{c} h^2  $ & $  0.1176\pm 0.0026 $ & $  0.1181\pm 0.0011 $ & $  0.1192\pm 0.0014 $ & $  0.11862\pm 0.00092 $ \\ 
$ 100\theta_\mathrm{MC}  $ & $  1.04109\pm 0.00047 $ & $  1.04101\pm 0.00037 $ & $  1.04080\pm 0.00026 $ & $  1.04087\pm 0.00024 $ \\ 
$ n_\mathrm{s}  $ & $  0.9682\pm 0.0083 $ & $  0.9665\pm 0.0048 $ & $  0.9646\pm 0.0048 $ & $  0.9664\pm 0.0040 $ \\ 
$ \log(10^{10} A_\mathrm{s})  $ & $  3.093\pm 0.044 $ & $  3.086\pm 0.023 $ & $  3.068\pm 0.027 $ & $  3.079\pm 0.021 $ \\ 
$ H_0  $ & $  68.2\pm 1.2 $ & $  67.93\pm 0.51 $ & $  67.49\pm 0.62 $ & $  67.74\pm 0.41 $ \\ 
$ \sigma_8  $ & $  0.824\pm 0.012 $ & $  0.8229\pm 0.0077 $ & $  0.8184\pm 0.0083 $ & $  0.8215\pm 0.0073 $ \\ 

\hline \hline
\end{tabular} }
\end{center}
\caption{Constraints at 68\% (95\%) CL on cosmological parameters obtained within the $\Lambda$CDM model of cosmology by different datasets involving temperature and polarization measurements from Planck likelihood \texttt{Camspec}.}
\label{tab.results.lcdm.camspec}
\end{table*}

\begin{table*}[!ht]
\begin{center}
\renewcommand{\arraystretch}{1.5}
\resizebox{\textwidth}{!}{
\begin{tabular}{l c c c c c c c c c c c c c c c }
\hline
\textbf{Parameter} & \textbf{\texttt{HiLLiPoP-TT+lensing} } & \textbf{ \texttt{HiLLiPoP-TT+lensing+low-z} } & \textbf{ \texttt{HiLLiPoP-TTTEEE+lensing} } & \textbf{  \texttt{HiLLiPoP-TTTEEE+lensing+low-z} } \\ 
\hline\hline

$ \tau_\mathrm{reio}  $ & $  0.076\pm 0.024\, ( 0.076^{+0.046}_{-0.046} ) $ & $  0.077\pm 0.012\, ( 0.077^{+0.025}_{-0.023} ) $ & $  0.073\pm 0.014\, ( 0.073^{+0.028}_{-0.028} ) $ & $  0.076\pm 0.011\, ( 0.076^{+0.023}_{-0.022} ) $ \\ 
$ \Omega_\mathrm{b} h^2  $ & $  0.02224\pm 0.00025 $ & $  0.02224\pm 0.00019 $ & $  0.02230\pm 0.00014 $ & $  0.02232\pm 0.00012 $ \\ 
$ \Omega_\mathrm{c} h^2  $ & $  0.1185\pm 0.0025 $ & $  0.1184\pm 0.0011 $ & $  0.1186\pm 0.0013 $ & $  0.11839\pm 0.00089 $ \\ 
$ 100\theta_\mathrm{MC}  $ & $  1.04097\pm 0.00048 $ & $  1.04099\pm 0.00038 $ & $  1.04082\pm 0.00027 $ & $  1.04085\pm 0.00025 $ \\ 
$ n_\mathrm{s}  $ & $  0.9660\pm 0.0076 $ & $  0.9662\pm 0.0044 $ & $  0.9679\pm 0.0044 $ & $  0.9686\pm 0.0034 $ \\ 
$ \log(10^{10} A_\mathrm{s})  $ & $  3.077\pm 0.042 $ & $  3.078\pm 0.022 $ & $  3.073\pm 0.025 $ & $  3.078\pm 0.020 $ \\ 
$ H_0  $ & $  67.8\pm 1.1 $ & $  67.81\pm 0.50 $ & $  67.76\pm 0.59 $ & $  67.85\pm 0.40 $ \\ 
$ \sigma_8  $ & $  0.820\pm 0.011 $ & $  0.8208\pm 0.0074 $ & $  0.8189\pm 0.0079 $ & $  0.8208\pm 0.0070 $ \\ 

\hline \hline
\end{tabular} }
\end{center}
\caption{Constraints at 68\% (95\%) CL on cosmological parameters obtained within the $\Lambda$CDM model of cosmology by different datasets involving temperature and polarization measurements from the Planck likelihood \texttt{HiLLiPoP}.}
\label{tab.results.lcdm.hillipop}
\end{table*}

Considering only small-scale temperature anisotropies along with low-redshift data, we obtain $\tau=0.079\pm0.012$ for \texttt{Camspec-TT+lensing+low-z} and $\tau=0.077\pm0.012$ for \texttt{HiLLiPoP-TT+lensing+low-z}. These results can be compared with those inferred from \texttt{plik-TT+lensing+low-z}, which yield $\tau=0.079\pm 0.013$. Apart from a marginal reduction in uncertainties, we note that the results remain consistent across all likelihoods. When we include small-scale polarization data, the values become $\tau=0.075\pm0.011$ for \texttt{Camspec-TTTEEE+lensing+low-z} and $\tau=0.076\pm0.011$ for \texttt{HiLLiPoP-TTTEEE+lensing+low-z}. Once again, these results demonstrate full consistency. They can be compared with the previously obtained results for \texttt{plik-TTTEEE+lensing+low-z}, which yield $\tau=0.080\pm0.012$. In this case, we can speculate that the improved treatment of small-scale polarization spectra leads to a slight shift towards values of $\tau\sim 0.075$ rather than $\tau\sim0.080$. However, such a shift in the mean value is statistically insignificant compared to the uncertainty and does not produce substantial differences in the conclusions we derived in the main paper.

Therefore, we conclude this appendix by emphasizing once again how small-scale temperature and polarization data, along with lensing observations and low-redshift probes, can provide independent constraints on the optical depth to reionization that do not rely on large-scale E-mode measurements. These results remain consistent across different experiments (Planck and ACT) and among different likelihoods (within the same experiment) that employ diverse techniques for handling foregrounds, such as \texttt{plik}, \texttt{Camspec}, and \texttt{HiLLiPoP}.

\end{widetext}

\clearpage

\bibliography{Bibliography}

\end{document}